\documentclass[preprint]{imsart}

\RequirePackage{amsthm,amsmath,amsfonts,amssymb}
\RequirePackage[authoryear]{natbib}
\RequirePackage[colorlinks,citecolor=blue,urlcolor=blue]{hyperref}
\RequirePackage{graphicx} 

\usepackage{array}
\usepackage{algorithm}
\usepackage{algorithmicx}
\usepackage[noend]{algpseudocode}
\usepackage{bm}
\usepackage{bbm}
\usepackage{bookmark}
\usepackage{booktabs}
\usepackage{dcolumn}
\usepackage{enumitem}
\usepackage{fancyhdr}
\usepackage[T1]{fontenc}
\usepackage{hyperref}
\usepackage[utf8]{inputenc}
\usepackage{lastpage}
\usepackage{mathrsfs}
\usepackage{mathtools}
\usepackage{multirow}
\usepackage[section]{placeins}
\usepackage{soul}
\usepackage{subcaption}
\usepackage{titlesec}
\usepackage{upgreek}
\usepackage{color}

\usepackage[margin=3.25cm]{geometry}

\newcounter{thm}[section]  

\startlocaldefs
\theoremstyle{plain}
\theoremstyle{remark}
\newtheorem{rem}[thm]{Remark}

\DeclareMathOperator*{\argmin}{arg\,min}

\DeclareMathOperator{\cov}{\mathrm{cov}}
\DeclareMathOperator{\diag}{\mathrm{diag}}
\DeclareMathOperator{\E}{\mathbb{E}}

\DeclareMathOperator{\var}{\mathrm{var}}

\newcommand*{\adj}{^{\mathsf{H}}}
\newcommand*{\trans}{^{\mkern-1.5mu\mathsf{T}}}

\DeclareMathOperator{\pr}{\uppi}

\renewcommand{\Re}{\mathcal{R}e}

\DeclareMathOperator{\I}{\mathbbm{1}}

\DeclareMathOperator{\Gaussian}{\mathcal{N}}
\DeclareMathOperator{\GP}{\mathcal{GP}}
\DeclareMathOperator{\Brain}{\mathcal{B}}

\bmdefine{\bLambda}{\bm{\Lambda}}
\bmdefine{\bSigma}{\bm{\Sigma}}
\bmdefine{\bTheta}{\bm{\Theta}}
\bmdefine{\bdelta}{\bm{\delta}}
\bmdefine{\blambda}{\bm{\lambda}}
\bmdefine{\bmu}{\bm{\mu}}
\bmdefine{\bphi}{\bm{\phi}}
\bmdefine{\btheta}{\bm{\theta}}
\bmdefine{\bxi}{\bm{\xi}}

\bmdefine{\bA}{\bm{A}}
\bmdefine{\bC}{\bm{C}}
\bmdefine{\bF}{\bm{F}}
\bmdefine{\bI}{\bm{I}}
\bmdefine{\bK}{\bm{K}}
\bmdefine{\bM}{\bm{M}}
\bmdefine{\bP}{\bm{P}}
\bmdefine{\bQ}{\bm{Q}}
\bmdefine{\bU}{\bm{U}}
\bmdefine{\bV}{\bm{V}}
\bmdefine{\bW}{\bm{W}}
\bmdefine{\bY}{\bm{Y}}

\bmdefine{\ba}{\bm{a}}
\bmdefine{\bc}{\bm{c}}
\bmdefine{\bd}{\bm{d}}
\bmdefine{\bp}{\bm{p}}
\bmdefine{\bq}{\bm{q}}
\bmdefine{\br}{\bm{r}}
\bmdefine{\bs}{\bm{s}}
\bmdefine{\bu}{\bm{u}}
\bmdefine{\bv}{\bm{v}}
\bmdefine{\bw}{\bm{w}}
\bmdefine{\bx}{\bm{x}}
\bmdefine{\bz}{\bm{z}}

\DeclareMathOperator{\vox}{\bm{\mathrm{v}}}

\DeclareMathOperator{\fft}{\mathcal{F}}
\DeclareMathOperator{\ifft}{\mathcal{F}^{-1}}

\endlocaldefs
\begin{document}

\begin{frontmatter}
\title{Bayesian Inference for Brain Activity from Functional Magnetic
  Resonance Imaging Collected at Two Spatial Resolutions}
\runtitle{Dual Resolution fMRI}

\begin{aug}
\author[A]{\fnms{Andrew S.}
  \snm{Whiteman}\ead[label=e1]{awhitem@umich.edu}},
\author[B]{\fnms{Andreas J.}
  \snm{Bartsch}\ead[label=e2,mark]{dr.bartsch@radiologie-bamberg.de}} 
\author[A]{\fnms{Jian}
  \snm{Kang}\ead[label=e3,mark]{jiankang@umich.edu}} 
\and
\author[A]{\fnms{Timothy D.}
  \snm{Johnson}\ead[label=e4,mark]{tdjtdj@umich.edu}} 
\address[A]{Department of Biostatistics, University of Michigan School
  of Public Health, \printead{e1} \printead{e3} \printead{e4}} 

\address[B]{Radiologie Bamberg and Department of Neuroradiology,
  University of Heidelberg, \printead{e2}} 
\end{aug}

\begin{abstract}
Neuroradiologists and neurosurgeons increasingly opt to use functional
magnetic resonance imaging (fMRI) to map functionally relevant brain
regions for noninvasive presurgical planning and intraoperative
neuronavigation. This application requires a high 
degree of spatial accuracy, but the fMRI signal-to-noise ratio (SNR)
decreases as spatial resolution increases. In practice, fMRI scans can
be collected at multiple spatial resolutions,  and it is of
interest to make more accurate inference on brain activity by
combining data with different resolutions. To this end, we
develop a new Bayesian model to leverage both better anatomical
precision in high resolution fMRI and higher SNR in
standard resolution fMRI. We assign a Gaussian process prior to
the mean intensity function and develop an efficient, scalable
posterior computation algorithm to integrate both sources of data.
We draw posterior samples using an algorithm analogous to
Riemann manifold Hamiltonian Monte Carlo in an expanded parameter
space. We illustrate our method in analysis of presurgical fMRI
data, and show in simulation that it infers the mean intensity more
accurately than alternatives that 
use either the high or standard resolution fMRI data alone.
\end{abstract}


\end{frontmatter}


\section{Introduction}

Neurosurgery presents a unique set of challenges to the operating
surgeon. Treatment of brain tumors, for example, is handled
primarily by surgical resection when possible. Gliomas are
often infiltrative, however, and as a result may be impossible to
remove entirely \cite[]{jovvcevska2013glioma, stippich2015clinical}. 
The neurosurgeon's goal is typically to resect as much of
the tumor as possible while avoiding damage to surrounding healthy 
areas of brain tissue, requiring precise structural and
functional information. Although the structure of the human brain
shares a gross organization common across individuals, functional
neuroanatomy may vary between patients and within regions
\cite[e.g.][]{large2016individual}, highlighting the need for
within-patient precision. Here we propose a model that leverages the
massive amount of spatial data available in individual functional
magnetic resonance imaging (fMRI) scans to help guide presurgical
planning by identifying functionally relevant brain regions in a
patient-specific manner.

Traditionally, electrocortical interference is used to map brain
functional organization during surgery
\cite[e.g.][]{cordella2013intraoperative}, but this procedure is
highly invasive, lengthens surgery duration, and cannot be
incorporated into presurgical planning
\cite[]{stippich2015clinical}. Clinicians can also opt
to use imaging methods to help inform patient-specific presurgical
planning and intraoperative neuronavigation
\cite[e.g.][]{archip2007non-rigid, 
  nimsky2006intraoperative, durnez2013alternative,
  silva2018challenges}. FMRI may be used, 
for example, to map patient-specific functional areas, but the
data come with an inherent trade off. Surgeons would like to
collect information that is spatially precise, but the fMRI
signal-to-noise ratio (SNR) decreases as spatial resolution increases,
potentially making functional mapping more difficult
\cite[]{bodurka2007mapping}. In practice, modern scanners are equipped
to handle a variety of image resolutions by modifying magnetic pulse
sequences, so radiologists are in principle able to collect any
combination of scans advantageous for presurgical planning.


Our motivating datasets come from two separate fMRI experiments in
which preoperative patients performed cognitive tasks chosen to
localize brain regions involved in language processing
(see sections \ref{sec:data-methods} and \ref{sec:data-analysis} for
details). 
Each individual patient was administered their task over two separate
scanning runs, collected at different spatial resolutions. Details
vary by patient, but in both instances one run was collected at
``standard'' spatial resolution with voxel (volumetric pixel)
dimensions measuring approximately $3 \times 3 \times 3$ mm$^3$, and
the other was collected at ``high'' spatial resolution with
approximately $2 \times 2 \times 2$ mm$^3$ voxels.
Raw image time series data were preprocessed using standard software
\cite[][]{jenkinson2012fsl, woolrich2001temporal} to yield statistical
parametric maps for each spatial resolution that summarized patients'
fMRI activation over time. In this
paper, we propose a new Bayesian model to integrate both sources of
data, leveraging the anatomical/spatial precision of high resolution
fMRI and the SNR of standard resolution fMRI for enhanced
within-patient precision.
The primary goal of our model is to
reduce spatial noise while making inferential statements identifying
functional regions at the highest resolution available. Conceptually,
we accomplish this goal by modeling the mean intensity function of
both data sources as a Gaussian process. Gaussian processes induce a
probability measure on a functional space with distribution
characterized by a mean and covariance function
\cite[][]{rasmussen2006gaussian}. 
Conditional on the covariance function hyperparameters, which we
estimate from data, we conduct fully Bayesian inference on the mean
function measured at voxel locations in the high spatial resolution
image.

In addition to spatial precision,
computational complexity is also a major concern since excessive
latency between preoperative scanning and a patient's actual surgery
is undesirable.
Computation with spatial Gaussian process models typically involve 
decomposition of an $n \times n$ matrix, where $n$ is the number of
spatial locations. 
Between the two image types there are over 200,000 unique spatial
locations in each of our motivating datasets, rendering usual
computational approaches to inference intractable in most computing
environments. Here, we outline a modification of the typical
Hamiltonian Monte Carlo (HMC) algorithm that makes this inference not
only feasible but computationally efficient. 
To do so, we propose a dual resolution mapping prior that generalizes
the existing Gaussian predictive process framework
\cite[e.g.][]{seeger2003fast, banerjee2008gaussian} to our setting
with multiple data sources. Our 
algorithm further harnesses a parameter expansion idea from
\cite{wood1994simulation} to sample from the posterior using Riemann
manifold Hamiltonian dynamics \cite[][]{girolami2011riemann} in an
ultrahigh dimensional parameter space. 

\begin{figure}[!ht]
\centering
\includegraphics[width=0.5\textwidth]{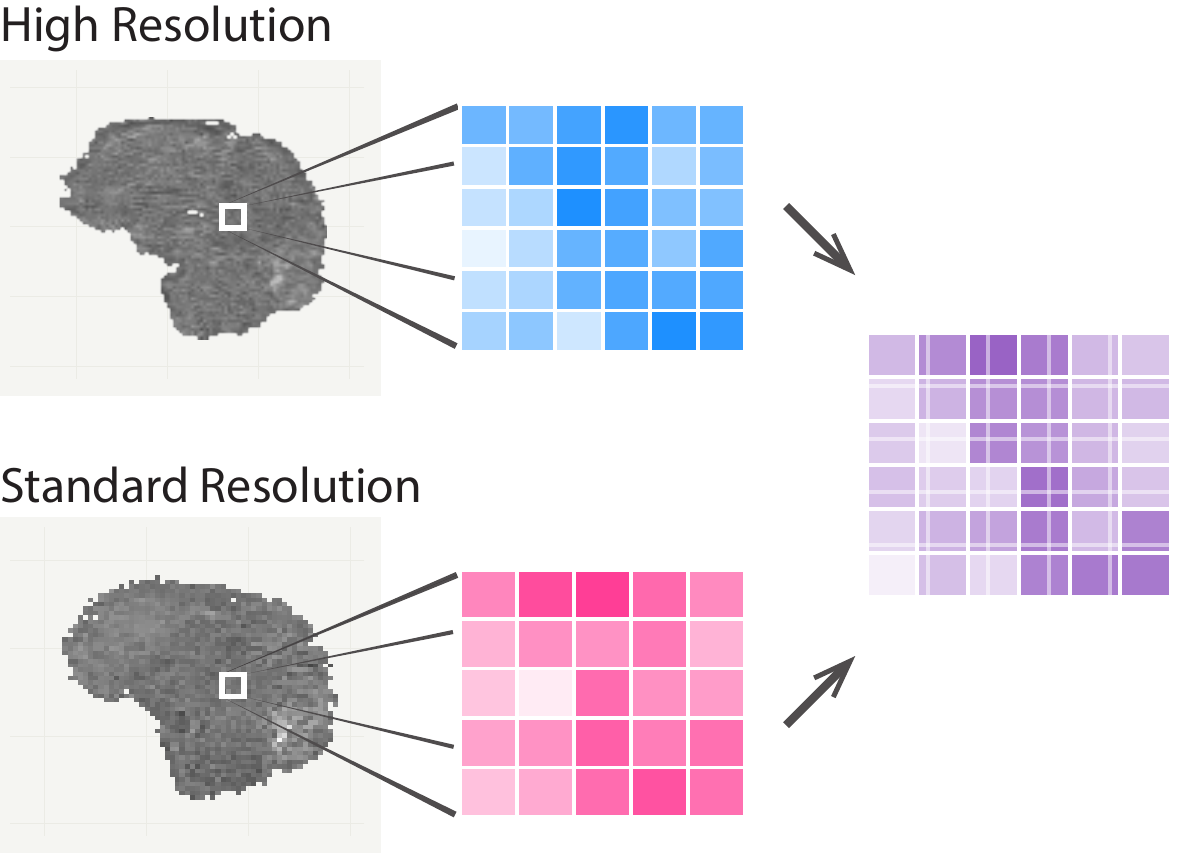}
\caption{Schematic of aims and difficulties with
  integration of fMRI data collected at multiple spatial
  resolutions. Images collected at different resolutions exhibit
  inherently different levels of noise. We would like to reduce
  spatial noise while making inferential statements at the highest
  resolution available, but voxel locations may not align in general.
  See the online version for color figures.
  \protect{\label{fig:dualres-problem}}}
\end{figure}

Our model is related to existing literature from the field of spatial
satistics that consider the ``change of support problem''
\cite[e.g.][]{gelfand2001change, fuentes2005model,
  berrocal2012space}. Such models have been used, for example, to
combine data from air pollution monitoring sites with simulations from
physical models for prediction at unobserved locations and model
validation. Studies such as these commonly model conditional
relationships between data sources, for example by regressing measured
air pollution onto physical model output. Our multi-resolution imaging
paradigm is related in the sense that we would like to use standard
resolution data to improve inference in high resolution space.
This goal, however, is complicated by
the fact that high and standard spatial resolution voxels in general
only partially overlap with their neighbors in their complementary
image (see Fig. \ref{fig:dualres-problem}).
We will, however, take a different
approach by modeling both sources of data as joint outcomes. Not only
does this approach perhaps make more conceptual sense for modeling
multiple image types, it permits flexible and natural reconfiguration
in response to real world challenges. For example, if only one fMRI
resolution or session is available presurgically, the missing data can
be removed from the joint outcome. Though we discuss our method
exclusively in a functional neuroimaging context, the method can
easily generalize to other imaging modalities or indeed 
to spatial data with mixed supports more broadly.

Whereas the inferential goal of most neuroimaging studies is to
identify activated or deactivated brain regions while controlling the
family-wise error rate, we take a somewhat different approach given
specific presurgical needs.
In a neurosurgical context, clinicians are typically
more concerned with inaccurate labeling of functionally important
tissue as unimportant. To this end, we adopt a decision theoretic rule
from previous work to control the ratio of false negative to false
positive errors \cite[]{liu2016pre, liu2019mixed, muller2006fdr}. We
show in simulation that our dual resolution method achieves 
good accuracy for realistic effect sizes. Specifically, our method
outperformed single spatial resolution alternatives in terms of both
false negative and false positive error rates when the number of
discoveries was fixed across methods. Software to fit the dual and 
single resolution models discussed in this paper to data stored in the 
NIfTI data standard \cite[][]{cox2004nifti} is available online
\cite[][]{whiteman2022code}.


The body of this paper contains descriptions of our motivating
clinical datasets in section \ref{sec:data-methods}, and a summary of
the method we propose to handle the unique challenges of those data in
section \ref{subsec:proposed-model}. In
sections \ref{subsec:methods:prior-approximation} and
\ref{subsec:methods:posterior-computation}, we elaborate on our
approach to enable precise estimation and computation in such a large 
parameter space. We discuss a strategy to conduct inference based on
weighted trade offs between false negative and false positive errors in
section \ref{subsec:results:posterior-inference}.
We quantify our method's performance against single resolution
alternative methods in section \ref{sec:simulations}.
Section \ref{sec:data-analysis} reports on analyses of real patient
data using our proposed method for dual resolution fMRI.
Finally, we present an overall evaluation of our contributions in
section \ref{sec:discussion}. 


\section{Data and methods}
\label{sec:data-methods}

We developed the method presented here to analyze single-patient
presurgical fMRI data collected at two spatial resolutions.
Our first motivating dataset comes from a 62 year old right handed
woman---``patient 1''---who presented difficulties with reading,
finding, and comprehending words. This patient was subsequently found
to have a tumor in her left middle and inferior temporal
gyrus. Prior to surgery, the patient was scanned while performing 
a reading task to map brain areas associated with reading non-final
embedded clause sentences and language processing. Scans were
collected in two separate runs: once at standard $3 \times 3 \times
3.45$ mm$^3$ resolution ($64 \times 64 \times 48$ grid), and once at
high $1.8 \times 1.8 \times 2.3$ mm$^3$ resolution
($120 \times 120 \times 62$ grid).

Our second motivating dataset comes from an 18 year old right handed
woman---``patient 2''---who presented after a general seizure and was
subsequently found to have a cavernoma in her left temporal lobe
(see Appendix \ref{app:cavernomas} for more detail).
For cavernomas in critical areas, presurgical
fMRI is considered one option---as with brain tumors---to
map brain function noninvasively for presurgical planning and
intraoperative neuronavigation. Patient 2 was also 
scanned prior to surgery while performing a language processing task.
Her standard resolution data were collected with slightly
smaller $3 \times 3 \times 3.3$ mm$^3$ voxels
($64 \times 64 \times 48$ grid), and her high resolution data with
$1.8 \times 1.8 \times 2.2$ mm$^3$ voxels
($120 \times 120 \times 62$ grid).
As in this patient, cavernomas typically cause profound
T2$^*$-weighted MR signal loss, with blooming into surrounding brain
tissue. Signal loss is caused by abrupt differences in magnetic
susceptibility in apposed tissues and is a common occurrence in
clinical fMRI (e.g. intratumoral hemorrhages can cause similar
dropout). We use this patient's data to illustrate our model's
capacity to recover an estimate of activation in areas of such fMRI
signal loss.

FMRI time series preprocessing without spatial smoothing was
performed prior to our analysis using FSL software
\cite[][]{jenkinson2012fsl} and the FEAT tool
\cite[][]{woolrich2001temporal}. As will become clear in
section \ref{subsec:proposed-model}, our model imposes smoothness on
the image mean function, and so we avoided smoothing the data during
preprocessing (beyond the small amount of unavoidable smoothing that
can occur when time series images from the two spatial resolutions
are motion corrected and co-registered with one another). Smoothing is
an otherwise ubiquitous step in typical fMRI pipelines, but
over smoothing is not desirable for presurgical planning applications
as it may reduce spatial precision by, for example, smearing
activation into adjacent areas when the smoothing kernel is too
wide. Data were corrected for motion and temporally high pass
filtered, and marginal linear models were fit to the time series data
at each voxel to create summary statistic maps of task-related
activation. 

Preprocessing resulted in one unsmoothed $z$-statistic contrast
image for each fMRI resolution that summarized task-related activation
over the course of each respective scan. We went on to use the
generated test statistic maps as outcome data in our subsequent
analysis, treating the images as noisy measures of true activation.
Although we may find it beneficial to include both spatial
and temporal data in our modeling framework in future work, the
present model only explicitly represents a spatial process. As such,
throughout the rest of this paper we will use ``high resolution,'' for
example, as a stand in for ``high \emph{spatial} resolution'' etc. In
the greater imaging community, however, ``resolution'' could in
general relate to frequency of either spatial or temporal data
collection, or both. We give additional details regarding
patient data collection and image preprocessing in Appendix
\ref{app:fmri}.


\subsection{Bayesian dual resolution mapping}
\label{subsec:proposed-model}

Let $\Brain$ denote a generic brain image space, and let $B_h
\subset \Brain$ and $B_s \subset \Brain$ denote the sets of
spatial locations in the brain where high and standard
resolution functional MRI data are collected, respectively. For
reference, the number of voxels $|B_h| \approx 200,000$ in the high
resolution image, and $|B_s| \approx 50,000$ in the standard
resolution image. Each atom $\vox \in \Brain$ is a three dimensional
vector of spatial coordinates relative to some origin
point $\vox_0 \in \Brain$; the Euclidean distance between any two
points, $\vox, \vox' \in \Brain$ can be represented $\lVert \vox
- \vox' \rVert_2$, and is typically measured in millimeters.
Although data at a given voxel is associated with a small volume, we
follow common practice and essentially treat that data as observed on
location $\vox$ exactly. 
In general, even voxels that overlap between the two image types may
not have the same centers, so that the set of points in the
intersection $B_h \cap B_s$ may be empty.

Conceptually, we motivate our proposed model as follows. Let
$Y_h(\vox_h)$ denote the high resolution imaging outcome at voxel
$\vox_h$, and let $Y_s(\vox_s)$ denote the standard resolution imaging
outcome at voxel $\vox_s$. For the same patient performing the same
cognitive task in the same scanner, we make the assumption that
$Y_h(\vox_h)$ and $Y_s(\vox_s)$ are realizations from a unifying
generative process. Let $\Gaussian(\mu, \sigma^2)$ denote a Gaussian
distribution with mean $\mu$ and variance $\sigma^2$. We model the
data as jointly Gaussian,
\begin{align}
  Y_h(\vox_h) &\sim \Gaussian(\mu(\vox_h), \sigma_h^2),
              \qquad \vox_h \in B_h \notag \\
  Y_s(\vox_s) &\sim \Gaussian(\mu(\vox_s), \sigma_s^2),
              \qquad \vox_s \in B_s
              \label{eqn:gp-likelihood}
\end{align}
where $\mu(\vox)$ represents the expected intensity of brain activity
in voxel $\vox \in \Brain$, and $\sigma_h^2$ and $\sigma_s^2$ are
noise variances in the high and standard resolution images,
respectively.
Because our data were not smoothed, we modeled noise as a
spatially independent and additive process. Given the known phenomenon
that SNR increases with voxel volume
\cite[e.g.][]{bodurka2007mapping}, we expect standard resolution
images to be less noisy than high resolution images. We therefore
adopted a weakly informative prior for the noise variances with the
restriction $\sigma_h^2 > \sigma_s^2$: 
\begin{equation}
  \label{eqn:noise-variances}
  \pr(\sigma_h^2, \sigma_s^2) \propto \sigma_h^{-2} \sigma_s^{-2}
  \I(0 < \sigma_s^2 < \sigma_h^2),
\end{equation}
where $\I(\cdot) \in \{0, 1\}$ is the event indicator function
($\I(\mathcal{A}) = 1$ if $\mathcal{A}$ occurs, and 0 otherwise).

For functional maps, we are primarily interested in making inferences
about the mean intensity function, $\mu(\cdot)$, to which we assign a
mean zero Gaussian process prior, 
\begin{equation}
\label{eqn:gp-prior-functional}
  \mu(\vox) \sim \GP(0, K(\vox, \vox')). \\
\end{equation}
In our formulation, the function $\mu(\cdot)$ captures all of the
correlation between voxels and between the two images; conditional on
$\mu(\cdot)$, $Y_h(\vox_h)$ and $Y_s(\vox_s)$ are mutually independent
across all $\vox_h \in B_h$ and $\vox_s \in B_s$. We implicitly
assume that the two brain images share a real-world coordinate system,
and that $\mu(\vox)$ is correlated with $\mu(\vox')$ if the distance
$\lVert \vox - \vox' \rVert_2$ is small. A variety of preprocessing
techniques have been developed to align 3D images and ensure the
former assumption holds with minimal error
\cite[e.g.][]{reuter2010highly,jenkinson2012fsl}.

Since anatomical precision is paramount in our application, we would like
to conduct inference on $\mu(\cdot)$ for all locations in $B_h$. To
facilitate this goal while simultaneously modeling the
cross correlation between $\mu(\cdot)$ evaluated on locations in $B_h$
and $B_s$, we introduce a nonstationary covariance function to map
between data sets. For any $\vox, \vox' \in \Brain$, let,
\begin{equation}
  \label{eqn:kernel}
  K(\vox, \vox') = \begin{cases}
    k(\vox, \vox') & \text{if } \vox' \in B_h \\
    w\trans(\vox) k(B_h, \vox') &
    \text{otherwise}, 
  \end{cases}
\end{equation}
where $w(\cdot)$ is a vector of weights in a finite basis (defined
below; chosen so that the covariance function is symmetric for all
$\vox, \vox' \in \Brain$), and
$k(\cdot, \cdot)$ is some positive definite function with
range $\mathbb{R}_{>0}$. In our application, we take
$k(\cdot, \cdot)$ to be the isotropic radial basis function,
\begin{equation}
  \label{eqn:rbf}
  k(\vox, \vox') = \tau^2 \exp(-\psi \lVert \vox - \vox'
  \rVert_2^\nu), \qquad \tau^2, \psi > 0, \quad \nu \in (0, 2],
\end{equation}
and extend the notation to apply to sets of locations so that
$k(B_h, \vox') = [k(\vox_h, \vox')]_{\vox_h \in B_h}$ is a vector in
$\mathbb{R}^{|B_h|}$.
In \eqref{eqn:rbf}, $\tau^2 > 0$ is the ``partial sill'' or marginal
prior variance of $\mu(\cdot)$, the decay parameter $\psi > 0$ defines
the correlation bandwidth, and $\nu \in (0, 2]$ is the kernel exponent or
smoothness parameter. We define the covariance parameters $\btheta =
(\tau^2, \psi, \nu)\trans$; $\psi$ and $\nu$, are
commonly fixed prior to analyses, but because of the abundance of spatial
data in even a single brain image, in practice we recommend estimating
these parameters from data (see section
\ref{subsec:analysis:kernel-estimation} for details).
The custom kernel in \eqref{eqn:kernel} was designed to approximate a
``Gaussian parent process'' \cite[][]{banerjee2008gaussian} with
isotropic radial basis covariance \eqref{eqn:rbf} everywhere in 
$\Brain$. We arrange our presentation here to make clear that we use
\eqref{eqn:kernel} directly, and obtain exact inference with the prior
\eqref{eqn:gp-prior-functional} specified in this way. 
A careful choice of the weight function $w(\cdot)$, moreover, can
render the problem more computationally tractable.

\subsection{Construction of the covariance weights}
\label{subsec:methods:prior-approximation}

By the definition of a Gaussian process, $\mu(\vox)$ and $\mu(\vox')$
are jointly multivariate Gaussian distributed for any distinct
locations $\vox$ and $\vox'$. As a result, Gaussian process models
promote natural and flexible predictions of values of $\mu(\cdot)$ at
unobserved locations. For arbitrary collections of locations
$U = \{ \vox_1, \ldots, \vox_n \} \subset \Brain$ and
$V = \{ \vox'_1, \ldots, \vox'_m \} \subset \Brain$, we define
$\mu(U) = [\mu(\vox_i)]_{i = 1}^n$ as a vector in $\mathbb{R}^n$;
$K(U, \vox) = [K(\vox_i, \vox)]_{i=1}^n$ as a vector in
$\mathbb{R}^n$; and $K(U, V) = [K(\vox_i, \vox'_j)]_{i,j = 1}^{n,m}$
as a matrix in $\mathbb{R}^{n \times m}$. If, for example, $V$ is a
set of observed locations, and 
$U$ are unobserved locations, then conditional on $\mu(V)$, $\mu(U)$
is multivariate Gaussian distributed with mean $K(U, V) K(V, V)^{-1}
\mu(V)$ and variance $K(U, U) - K(U, V) K(V, V)^{-1} K(V, U)$.

Since imaging data is collected on a dense grid we often have no
need to predict outcomes at unobserved or non-brain locations, except
in cases of signal loss or other artifact.
We made use of the kriging or conditional distribution
relationships above primarily to define the basis weight function
$w(\cdot)$ to integrate information from both high and
standard resolution images.
We constructed the basis weights in \eqref{eqn:kernel} so that
$w(\vox) \approx K(B_h, B_h)^{-1} k(B_h, \vox)$, with the
``approximate'' relation explained below.
This formulation allowed us to leverage the relationship that
$w\trans(\vox) \bmu_h$ approximates the prior conditional expectation
of $\mu(\vox)$ given $\bmu_h = \mu(B_h)$.
As such, our construction in \eqref{eqn:kernel} generalizes a Gaussian
predictive process framework 
\cite[e.g.][]{seeger2003fast, banerjee2008gaussian}
to our setting with multiple data sources by using $B_h$ as a
high-dimensional reference set.
In general, within this framework we could have defined the weights 
$w(\cdot)$ based on any arbitrary set of knot locations
$B_* \subset \Brain$. Since inference at a fine spatial scale
typically requires a dense set of knot locations
\cite[e.g.][]{stein2007spatial}, we preferred to 
define $w(\cdot)$ based on all of $B_h$.

Our covariance function in \eqref{eqn:kernel} effectively employs
kriging methods to 
map $\mu(B_h)$ onto the locations in $B_s$ so that the standard
resolution data can still inform $\mu(B_h)$ in the posterior. 
Switching to vector notation, let $\bmu_s = \mu(B_s)$.
We express the prior in \eqref{eqn:gp-prior-functional},
\begin{equation}\label{eqn:approximate-prior}
  \pr(\bmu_h, \bmu_s) = \Gaussian\left( \bm{0},
    \begin{bmatrix}
      \bK_h & \bK_{h,s} \\
      \bK_{s,h} & \bK_{s,h} \bK_h^{-1} \bK_{h,s}
    \end{bmatrix} \right).
\end{equation}
where $\bmu_h$ and $\bmu_s$ are the means of the high, and standard
resolution images, respectively; we denote the marginal prior
variance of $\bmu_h$ by $\bK_h = K(B_h, B_h)$, the prior
covariance of $\bmu_h$ and 
$\bmu_s$ by $\bK_{h,s} = K(B_h, B_s)$, etc.
The obvious difficulty working with \eqref{eqn:approximate-prior}
directly is that the covariance matrix is large and dense and we need
to be able to compute its inverse in order to evaluate the prior. We
would like to make inferential statements about $\bmu_h$, but the
dimension of the submatrix $\bK_h$ ($n_h \approx$ 200,000) alone is
prohibitive on most hardware architectures---such a matrix would
require over $(1.8 \times 10^5)^2 \times 32 = 129.6$ Gb of memory just
to store in a single precision floating point format. Though the
memory requirement could be reduced by storing just the upper or lower
triangle, to sample $\bmu_h$ Cholesky decomposition of $\bK_h$ would
still require $\approx 1.9 \times 10^{15}$ floating point operations
(FLOPs) to compute.

In \eqref{eqn:approximate-prior}, the
covariance matrix has rank of at most $n_h$, and the implied
conditional density $\pr(\bmu_s \mid \bmu_h)$ is degenerate on $\bK_{s,h}
\bK_h^{-1} \bmu_h$.
Additionally, we represent the product $\bK_h^{-1} \bK_{h,s}$ by
matrix $\bW\trans = [w(\vox_s)]_{\vox_s \in B_s}$.
To induce sparsity and save computational resources, we defined $\bW$
in terms of neighborhoods of voxels in $B_h$.
For any $\vox \in \Brain$, let $N_h(\vox)$ denote a set of locations
in $B_h$ in an $r$-neighborhood of location $\vox$,
$N_h(\vox) = \{ \vox_h \in B_h : \lVert \vox_h - \vox \rVert_2 \leq r
\}$.
If $N_h(\vox)$ is empty, then we defined $w(\vox) = \bm{0}$; otherwise let
$\bK_{N_h(\vox)} = [k(\vox_i, \vox_j)]_{\vox_i, \vox_j \in N_h(\vox)}$,
let $\bm{k}_{N_h(\vox)} = [k(\vox_i, \vox)]_{\vox_i \in N_h(\vox)}$,
and let $\tilde{\bw} = \bK_{N_h(\vox)}^{-1} \bm{k}_{N_h(\vox)}$
denote a vector with implicit dependence on $N_h(\vox)$ where each
element corresponds with one location in 
$N_h(\vox)$. For non-empty $N_h(\vox)$, each element of $w(\vox)$
similarly corresponds with one location in $B_h$. We defined those
elements to be,
\begin{equation}\label{eqn:basis-weights}
  w_i(\vox) = \begin{cases}
    \tilde{w}_j & \text{if the $j^{th}$ location in $N_h(\vox)$
      corresponds to the $i^{th}$ location in $B_h$} \\
    0 & \text{otherwise}.
    \end{cases}
\end{equation}
With $\bW\trans = [w(\vox_s)]_{\vox_s \in B_s}$, the product
$\bW \bmu_h$ can be interpreted as a local kriging approximation of
$\bmu_s$ conditional on $\bmu_h$. Our definition of $\bW$ is
conceptually somewhat inspired by work on
Nearest Neighbor Gaussian Processes by
\cite{datta2016hierarchical,finley2019efficient}. A sensitivity
analysis over choice of $r$ is available in the Supplementary Material
\cite[][]{whiteman2022supp}.

The matrix $\bW$ can be entirely precomputed given the kernel
parameters, $\psi$, $\nu$, and a neighborhood radius, $r$. Equipped
with the matrix $\bW$, samples from
\eqref{eqn:approximate-prior} can be drawn by first sampling
$\bmu_h \sim \Gaussian(\bm{0}, \bK_h)$, and then computing
$\bmu_s = \bW \bmu_h$. 
In practice we treat $r$ as a hyperparameter and
condition analyses on it. In our data example
(see section \ref{sec:data-analysis}) we took the radius $r$ to be
roughly one FWHM length based on estimated prior
covariance and hyperparameters $\btheta$ (section
\ref{subsec:analysis:kernel-estimation}). This choice was motivated by the
desire to keep $r$ roughly in line with the width of 
\eqref{eqn:rbf} while keeping $\bW$
only modestly expensive to compute: for this choice of $r$, typical
neighborhood sizes $|N_h(\vox)|$ were on the order of 300--700
voxels in patient data. We next outline an efficient 
posterior computation algorithm for $\bmu_h$.


\subsection{Posterior computation}
\label{subsec:methods:posterior-computation}

\begin{figure}[!htb]
  \centering
  \begin{tabular}{ >{\centering\arraybackslash} m{0.28\textwidth}
    >{\centering\arraybackslash} m{0.28\textwidth} 
    >{\centering\arraybackslash} m{0.28\textwidth} }
    $\begin{bmatrix}
      \bm{10} & \bm{6}  & \bm{2}  & \bm{0}  & 2  & 6  \\
      \bm{6}  & \bm{10} & \bm{6}  & \bm{2}  & 0  & 2  \\
      \bm{2}  & \bm{6}  & \bm{10} & \bm{6}  & 2  & 0  \\
      \bm{0}  & \bm{2}  & \bm{6}  & \bm{10} & 6  & 2  \\
      2  & 0  & 2  & 6  & 10 & 6  \\
      6  & 2  & 0  & 2  & 6  & 10 \\
    \end{bmatrix}$  &
    \includegraphics[width=0.25\textwidth]{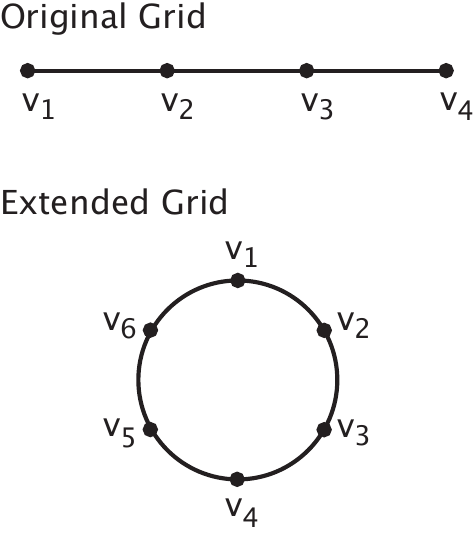}
                        &
    $\begin{bmatrix}
      C_0    & C_1    & \cdots & C_{m-1} \\
      C_{m-1} & C_0    & \cdots & C_{m-2}    \\
      \vdots & \vdots & \ddots & \vdots \\
      C_1    & C_2    & \cdots & C_0    \\
    \end{bmatrix}$
  \end{tabular}
  \caption{Example circulant matrix embedding.
    The left-most panel shows an example $4 \times 4$ Toeplitz matrix
    (bold) embedded within a $6 \times 6$ circulant matrix. In this
    simple example, the inner Toeplitz matrix might correspond with
    locations on a 1D grid (center panel). Conceptually, the outer
    circulant matrix can be taken to correspond with an extended grid,
    where an extended set of vertices have been ``wrapped around'' a
    circle. In the more general case (right-most panel), blocks $C_i$
    of a circulant-family matrix have symmetry such that 
    $C_{m-i} \equiv C_i$.
    \label{fig:circulant-embedding}
  }
\end{figure}

To facilitate computation we embedded the random field $\bmu_h$ within
a much larger random field, which we will call $\bu$. Our goal in
doing so was to be able to replace expensive matrix operations with
computations involving discrete Fourier transformations (DFTs) as we
show below. Conceptually, the augmented parameter space we chose can
be viewed to correspond with an extended grid of locations with
toroidal geometry.
In our data application, the resulting extended grid has about
$8.4 \times 10^6$ elements (grid dimensions
$256 \times 256 \times 128$). We treat this extended grid as if it
were a part of $B_h$ in the prior, with the result that the covariance
of the field $\bu$ has a nested block-circulant structure.
In the discussion to follow, we will use $\bC$ to denote the prior
variance of $\bu$. 
With this construction, we have added a large number of
auxiliary parameters, but have not changed the effective prior on
$\bmu_h$: the matrix $\bK_h$ is a principal submatrix of
$\bC$. Any Toeplitz-family matrix can be embedded in a
larger circulant-family matrix in this way.
For additional exposition, Fig. \ref{fig:circulant-embedding} shows a
simple example of this type of circulant embedding. In the figure,
the left and center panels illustrate circulant embedding for a 1D
grid. If working on a 2D grid, then schematically each block $C_i$ in
the right-most panel of Fig. \ref{fig:circulant-embedding} will be a
circulant matrix; on a 3D grid each block will itself be
block-circulant, etc.
This construction can be used to enable efficient simulation of random
Gaussian fields over dense grids as others have shown
\cite[e.g.][]{wood1994simulation, rue2005gaussian}
and as we summarize below.

Circulant matrix--vector products can be computed efficiently
with DFT software. Given the first row or column $\ba$---the so called
base---of a circulant matrix $\bA$ the product $\bA \bx$ can be
expressed as the discrete convolution $\ba \ast \bx$. Equivalently, by
the discrete convolution theorem, $\text{DFT}(\bA \bx) =
\text{DFT}(\ba) \odot \text{DFT}(\bx)$, where $\odot$ denotes an
elementwise or Hadamard product. The principle is the same with a
nested block-circulant matrix like $\bC$: the matrix has a base,
$\bc$, that is efficient to work with using 3D DFTs. In the present
case, $\bc$ can be precomputed (see the Supplementary Material) 
from the original grid dimensions
and covariance function $K(\cdot, \cdot)$. As
with any circulant matrix, $\bC$ can be diagonalized by two Fourier
matrices. If $\bF$ denotes a scaled 3D DFT matrix, and $\bF\adj$ its
adjugate, $\bF\adj \bC \bF = \diag(\blambda)$, where $\blambda$ are
the (complex) eigenvalues of $\bC$. With only the base $\bc$ in
memory, $\blambda = \fft(\bc) / N$ can be computed directly, where
$N$ is the number of elements in $\bc$, and $\fft(\cdot)$ denotes the
3D discrete Fourier transform.
We provide a simple algorithm to construct $\bc$ for any dense 3D grid
in the Supplementary Material \cite[][]{whiteman2022supp}.

\cite{wood1994simulation} took advantage of this relationship to
propose an efficient algorithm for simulation of random Gaussian
fields when the covariance of the field can be embedded within a
circulant matrix. 
For example, in our setting, we could sample from the prior
$\pr(\bmu_h \mid \btheta)$ by first 
drawing $\bz \sim \mathcal{CN}(\bm{0}, \bI)$, where
$\mathcal{CN}(0, v)$ denotes the circularly symmetric complex normal
distribution with variance $2v$.
Let $\ifft(\cdot)$ denote the 3D inverse DFT, let
$\ba^{\circ b} = [a_i^b]$ denote element-wise or Hadamard
exponentiation, and let $\Re(\ba)$ extract the real part of a complex
vector $\ba$. 
With $\blambda$ computed as above, we could then set
$\bu \gets \Re[ \fft\{ \blambda^{\circ 1/2} \odot \ifft(\bz) \} ]$,
and obtain a prior sample of $\bmu_h$ by simply discarding extraneous
elements of $\bu$.

There is no direct extension of the above \cite{wood1994simulation}
algorithm for posterior simulation in our setting. In part, this is
because we use different noise variance terms for our two data sources
\eqref{eqn:noise-variances}. Unless the diagonal noise terms
are exactly equal, the joint posterior variance of
$(\bmu_h\trans, \bmu_s\trans)\trans$ will not be Toeplitz in
general. We still draw inspiration from the work of
\cite{wood1994simulation}, however, and use the circulant matrix
relationships above to write an efficient Hamiltonian Monte Carlo
algorithm for posterior inference.
Details of this algorithm are presented in the Supplementary Material
\cite[][]{whiteman2022supp}, 
but the key components are: (i) as discussed, we embed $\bmu_h$ in a
higher dimensional random Gaussian field with a circulant covariance
matrix; and (ii) we construct a circulant ``mass matrix'' for our
HMC. Modification (i) allows us to be able to evaluate the log prior
and compute its gradient, and modification (ii) dramatically improves
mixing of the HMC chains.
As a result, our algorithm reduces the computational requirement
to evaluate the log prior on $\bmu_h$ roughly to $<$ 0.01 Gb and
$\approx 2 \times 10^9$ FLOPs. We now turn to remark on how we
summarize inference from our model in practice.

\subsection{Functional region detection}
\label{subsec:results:posterior-inference}
FMRI detects functionally relevant brain regions by recording
changes in oxygenated blood flow (BOLD signal). In a typical study,
practitioners identify these regions by thresholding voxelwise
statistical summaries in a manner that controls the false discovery
rate \cite[e.g.][]{genovese2002thresholding}. For presurgical
applications, it is at least as important to limit false negative
reports, since errors of this kind may potentially lead to damage
of healthy tissue. To this end, we adapted a decision theoretic
approach following previous work \cite[][]{muller2006fdr, liu2016pre,
  liu2019mixed}. We consider the loss function, 
\begin{multline}\label{eqn:loss-function}
    L(\bm{m}, \bdelta) = \sum_i {-f(m_i) \delta_i - \{1 - f(m_i)\}(1 -
      \delta_i)} \\
    + {k_1 f(m_i) (1 - \delta_i) + k_2 \{1 - f(m_i)\}
      \delta_i} + t \delta_i, 
\end{multline}
where $(k_1, k_2, t)$ are tunable constants, the
$m_i = | \mu_{h,i} | / \sqrt{\var( \mu_{h,i} )}$
are posterior $t$-statistic analogs measuring pointwise signal
strength in $\bmu_h$, and the $\delta_i \in \{0, 1\}$ are
pointwise statistical decisions (i.e. $\delta_i = 1$ reports a finding
at voxel $i$, and $\delta_i = 0$ otherwise). The function $f(\cdot)$
can be any monotonically increasing function restricted to $[0, 1]$,
and is intended to act as a proxy for
$\pr(\delta_i = 1 \mid \bY_h, \bY_s, \btheta, r)$.
Again, following previous work \cite[][]{liu2016pre, liu2019mixed}, we
take $f(m) = m / M$, where $M = \max_i \{m_i\}$. 

The loss function \eqref{eqn:loss-function} is composed of five terms,
each with a distinct importance: $-\sum_i f(m_i) \delta_i$ and
$-\sum_i \{1 - f(m_i)\} (1 - \delta_i)$ induce gains for correct
discoveries and correct non discoveries, respectively;
$k_1 \sum_i f(m_i) (1 - \delta_i)$ penalizes false negative errors;
$k_2 \sum_i \{1 - f(m_i)\} \delta_i$ penalizes false positive errors;
and $t \sum_i \delta_i$ penalizes the total number of
discoveries. Optimal decisions $\delta_i^*$ minimize the posterior
risk and follow,  
\begin{equation}\label{eqn:decision-rule}
\delta_i^* = \I\{ \bar{f}_i \geq (1 + k_2 + t) / (2 + k_1 + k_2) \},
\end{equation}
where $\bar{f}_i$ is the posterior expectation
$\E\{ f(m_i) \mid \bY_h, \bY_s, \btheta, r \}$, and the parameters
$(k_1, k_2, t)$ suggest a threshold based on a 
trade off between false negative and false positive errors.

Thresholds can be tuned with domain expert guidance
and/or varied dynamically, as a single static threshold may not be
sufficient for a surgeon's needs \cite[e.g.][]{stippich2015clinical}.
As a practical note, setting $k_2 = t = 1$ and varying
$k_1$ over the range $[5, 12]$ can provide good guidance, with
$k_1 = 7$ a reasonable default.
In one of our patient data analyses (below), we
set $t = 1$, $k_1 = 12$, and $k_2 = 1$ for inference.
As per our coauthor and collaborating
neuroradiologist's advice, this tuning parameter choice penalizes false
negative errors 12 times more heavily than false positive errors.
The other patient in our data was somewhat younger and less ill, with
no interictal speech or language impairments.
Consequentially her $z$-statistic images appeared to have a better
signal to noise ratio. For this patient, the suggestion was to set
$k_1 = 7$ and use a seven fold penalty ratio (not shown). In both
cases, the corresponding activation thresholds were confirmed visually
by comparison with results from intraoperative electrocortical
interference mapping.

\section{Simulation studies}
\label{sec:simulations}

We quantified the advantages of our proposed method with easier to
visualize simulations in two dimensions.
Our goal in simulation
was to evaluate how well the proposed model and alternative methods
recovered activation patterns in data. Typical fMRI studies use
significance testing as a means to identify functionally
relevant brain regions. To mimic this setting,
our simulation designs considered active regions embedded within low
variance signal (see Fig. \ref{fig:2d-design}).
As we discuss below, we further tried to mimic the patient
data by roughly matching simulated spatial signal smoothness and
signal-to-noise ratios to the real data.

\begin{figure}[!ht]
\centering
\begin{tabular}{ c c c c c c }
\multicolumn{3}{c}{High Resolution} & & \multicolumn{2}{c}{Standard Resolution} \\
\textsf{Active Regions} & \textsf{Mean Intensity} & \textsf{Data} & & \textsf{Mean Intensity} & \textsf{Data} \\
\includegraphics[width=0.16\textwidth]{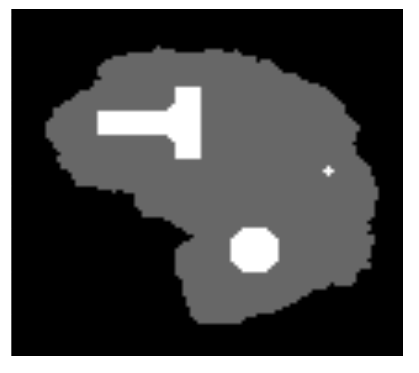} & 
\includegraphics[width=0.16\textwidth]{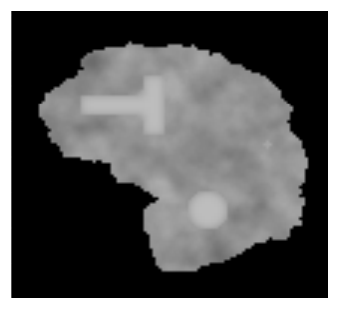} & 
\includegraphics[width=0.16\textwidth]{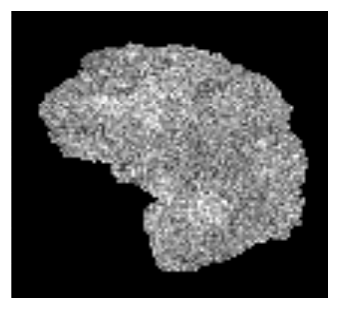}
                        & &
\includegraphics[width=0.16\textwidth]{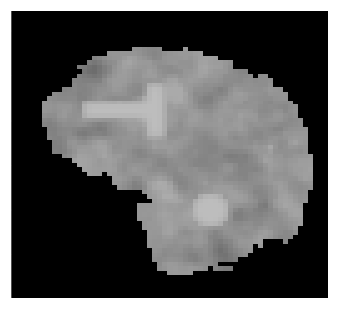} & 
\includegraphics[width=0.16\textwidth]{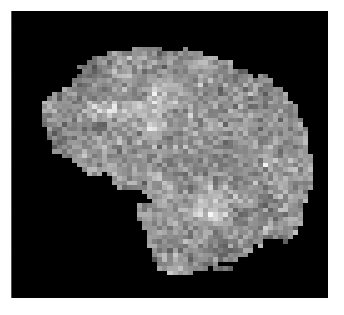}
\end{tabular}
\caption{Simulation design example with SNR$_h$ = 0.1 and
  SNR$_s$ = 0.2. Non-activation smooth signal has marginal variance
  0.2 and 6 mm FWHM Exponential correlation; activation signal has
  mean 2.
  \protect{\label{fig:2d-design}}} 
\end{figure}


\subsection{Simulations on 2D grids}
\label{subsec:simulations:2d}

Figure \ref{fig:2d-design} illustrates
our general approach to data simulation. In the figure, active regions
were drawn in a midsagittal plane including a T-shaped region, a
circular region, and a four voxel square.
These signals were
created by smoothing binary images with a six millimeter full width at
half maximum (FWHM) Gaussian kernel, scaling by a factor of two, and
thresholding the result at 0.4. We then embedded the active regions
within random draws from a 2D random field with mean zero,
marginal variance 0.2, and 6 mm FWHM Exponential or Gaussian
correlation functions. We treated the resultant images as true nonzero
mean intensity images, with ``active'' voxels given only by the
smoothed T, circle, and square shapes;
on average about 11\% of activation-adjacent voxels
would have had signal strength within $\pm 0.25$ standard deviations
of their active neighbors.
As per the patient data, we treated voxels in this plane as
$(1.8 \times 1.8)$ mm for high resolution (4,722 voxels total), and as
$(3 \times 3)$ mm for standard resolution (1,853 voxels). With our
design, there were exactly 450 active voxels in the high resolution
slice ($9.5\%$; see Fig. \ref{fig:2d-design}).

We adopted this method to generate ``high resolution'' mean images, or
$\bmu_h$ as in section \ref{subsec:methods:posterior-computation},
and projected $\bmu_h$ into ``standard resolution'' space by
multiplying by $\bW$ as in section
\ref{subsec:methods:prior-approximation} to generate corresponding
standard resolution mean images. In all simulation settings, $\bW$ was
constructed using the true 6 mm FWHM Exponential or Gaussian
background signal correlation functions, and an extent radius $r$ defined as the
distance after which the correlation would drop below 0.05. To
simulate observed outcome data, we added independent
Gaussian noise to the mean intensity images, modulated the noise
variances to control SNRs of the simulated high and standard
resolution images, and ran 100 replicates per parameter
combination. We took the SNR to be the ratio of the second 
moment of the mean to the variance of the noise, and set this to be
one of $\{0.1, 0.2\}$ for high resolution images (SNR$_h$). We
parameterized standard resolution noise in terms of the ratio of
standard to high resolution SNR (SNR$_s$:SNR$_h$), and set this ratio
to one of $\{1, 2, 4\}$. In the first case, the standard resolution
image would not provide additional signal-to-noise support as it
typically would in real data. We considered this a worst case
scenario. The latter two settings were chosen so that the standard
resolution image provided increasingly large signal-to-noise support,
where we expected our dual resolution method to dominate. In our 
analysis of the patient 1 data, based on the fits of our high and
standard single resolution alternative models, we estimated SNR$_h$
$\approx 0.18$ and SNR$_s$ $\approx 0.44$ based on their posterior
means (ratio SNR$_s$:SNR$_h$ $\approx$ 2.4).

\subsection{Recovery of simulated activation regions in 2D images}
\label{subsec:simulations:2d-recovery}

\begin{table}[ht]
\centering
\begin{tabular}{ l l c c c l }
  \multicolumn{1}{c}{\emph{Model}} 
  &  \multicolumn{1}{c}{\emph{Kernel}} 
  & \emph{SNR}$_s$\emph{:SNR}$_h$
  & \emph{SNR}$_h$
  & \emph{MSE}
  & \multicolumn{1}{c}{\emph{False --}} \\ 
  \hline
\textbf{Dual} & Exponential & 1 & 0.1 & \textbf{0.20} & \textbf{31.8\% (0.4)} \\ 
  High & Exponential & 1 & 0.1 & 0.23 & 34.0\% (0.5) \\ 
  Naive & Exponential & 1 & 0.1 & 0.30 & 43.6\% (0.4) \\ 
  Std & Exponential & 1 & 0.1 & 0.47 & 43.1\% (0.6) \\ 
  \textbf{Dual} & Exponential & 2 & 0.1 & \textbf{0.18} & \textbf{30.6\% (0.4)} \\ 
  High & Exponential & 2 & 0.1 & 0.23 & 34.0\% (0.5) \\ 
  Naive & Exponential & 2 & 0.1 & 0.29 & 42.7\% (0.4) \\ 
  Std & Exponential & 2 & 0.1 & 0.43 &  40.6\% (0.4) \\
\end{tabular}
\caption{Selected results for estimation and inference
  quality in 2D simulations. Results for the \emph{High} 
  resolution method do not change across the different SNR ratios,
  but are repeated to facilitate comparison. \emph{Model} denotes the
  image combination used in the analysis, and \emph{Kernel} gives the
  correlation pattern of low variance background signal. \emph{MSE}
  refers to mean squared error computed over the entire high
  resolution mean parameter vector; the simulation standard error of
  this metric was on the order of $10^{-3}$ for all simulation
  settings and so was omitted for brevity. \emph{False --} reports the
  mean (SE) false negative error rate when the number of discoveries
  was fixed at 450. One hundred replicates per parameter combination;
  additional results with different kernel and SNR$_h$ parameter
  settings are summarized in the Supplementary Material.
  \protect{\label{tab:2d-sims-main}}}
\end{table}

In each simulation, models were conditioned on the true
$\btheta = (\tau^2, \psi, \nu)\trans$ used to generate the low 
variance mean fields. We chose to condition on the true $\btheta$ so
as to explicitly focus our simulation results on estimation of and
inference on the image mean intensities.
We compared performance of our dual resolution model
\eqref{eqn:gp-likelihood} against single resolution alternative
methods:
(i) a related Gaussian process model that only considered the high
resolution data, (ii) the same model but considering only standard
resolution data (kriging the posterior mean of $\bmu_s$ to the
locations in $B_h$), and (iii) a method that we term naive data
averaging. For the alternative high and
standard resolution models, we used a Gaussian process to model the
the mean of the data as in \eqref{eqn:gp-prior-functional}. 
For the naive alternative, we
estimated the matrix $\bW$ (defined in section 
\ref{subsec:methods:prior-approximation}) from the data and used it to 
interpolate standard resolution data into the high resolution
space. We then treated a simple pointwise average of the high and
interpolated standard resolution images---i.e. $\bar{\bY}_{hs} =
(\bY_h + \bW\trans \bY_s) / 2$---as data in the alternative high
resolution model (i). This approach is conceptually similar to
previous work in this area \cite[][]{liu2019mixed}. The high
resolution method (i) served as our primary comparison point both
because of its inherent spatial resolution and because it tended to be
the best competing method in our simulations (see section
\ref{sec:simulations}).

Table \ref{tab:2d-sims-main} presents selected results for estimation
and inference quality in our 2D simulation settings. Results are
presented predominantly for the setting with SNR$_h$ = 0.1,
the SNR$_s$:SNR$_h$ ratio set to two, and an Exponential correlation
function to roughly approximate our patient data (also reflected
in Fig. \ref{fig:2d-design}). We provide a comparison
point with the SNR ratio equal to one for additional interest. More
extensive results are available in the Supplementary Material
\cite[][]{whiteman2022supp}. 
In the table, \emph{MSE} denotes the mean squared error of the estimated
$\bmu_h$, computed over pixels in our simulated high resolution
slices. We treat MSE as a measure of estimation quality, and
report that in all simulation settings considered, MSE was lowest
for the dual resolution models. This result indicates that when the
same mean intensity function underlies both high and standard
resolution images and the kernel function is estimated accurately, the
model that used joint information from both imaging modalities outperformed
possible single resolution alternatives. Interestingly, when the
background intensity was generated with an Exponential kernel, as in
Table \ref{tab:2d-sims-main}, the model that used only high resolution
data was the second best performer, underscoring the importance of
spatial precision in estimation.

We also report false negative rates---the measure of inference we are
most concerned with in our framework---for each model in Table
\ref{tab:2d-sims-main}. In the table, we set parameters $k_1$, $k_2$,
and $t$ in our decision rule \eqref{eqn:decision-rule} independently for
each model type so as to control the total number of discoveries to
exactly 450 (the same as the number of pixels we considered truly
active in the simulations). The actual decisions corresponding
to these thresholds are shown in Fig. \ref{fig:2d-inference}
(\emph{right}) for a single representative simulation iteration. We
emphasize that thresholds here were chosen as an objective point of comparison
across the alternative methods, not by optimizing any kind of
inferential criteria. In Fig. \ref{fig:2d-inference} (\emph{left}), 
we show that the dual resolution model would give superior inference
for any set of decision rule thresholds that fix the false negative
rate at a single value across all methods.

\begin{figure}[hbt]
\centering
\includegraphics[height=0.4\textwidth]{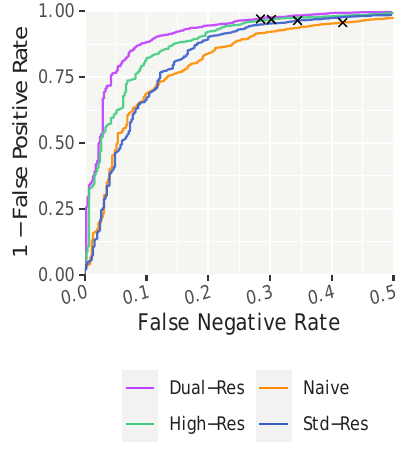}\hspace{0.05\textwidth} 
\raisebox{0.05\height}{\includegraphics[height=0.4\textwidth]{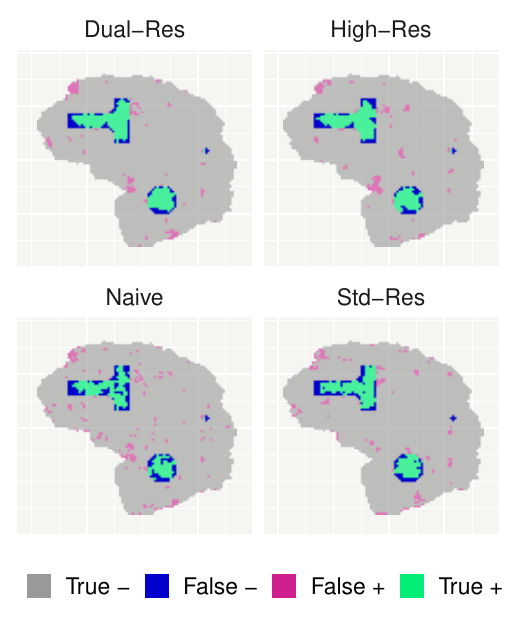}} 
\caption{Inference quality in 2D simulations. (\emph{Left})
  Receiver operating characteristic (ROC) curves comparing dual and
  single resolution methods to a naive data averaging approach in a
  setting that matches the data in Fig. \ref{fig:2d-design}. The
  curves show that for almost any given false negative rate, the dual
  resolution method can have a uniformly lower false positive rate
  than alternative single resolution methods. The $\times$'s mark the
  thresholds used to generate the inferential summary on the
  (\emph{right}). These thresholds limit the total number of
  discoveries to 450 across all four
  methods. \protect{\label{fig:2d-inference}}} 
\end{figure}

\section{Patient data analysis}
\label{sec:data-analysis}

As noted, our first motivating dataset---``patient 1''---comes from a
right handed 62 year old woman who presented primarily with
difficulties reading (pure alexia). This patient was found to have a
large tumor in her left middle and inferior temporal gyrus; following
partial surgical resection, the tumor was classified as a glioblastoma
multiforme.
Our second motivating dataset---``patient 2''---comes from an 18 year
old right handed woman who presented after a general seizure. This
patient was found to have a relatively large cavernoma adjacent to
insular cortex and the transverse temporal gyrus.
Both patients were scanned prior to surgery while performing
a reading task with a 30 
second on/off block design to map brain areas associated with 
reading and subsequent language processing. The task consisted of
silent reading in interleaved blocks of non-final embedded clause
sentences (on; eight blocks) and strings of consonants (control; eight
blocks).

Details of our fMRI acquisition protocol and preprocessing are given
in Appendix \ref{app:fmri}. Preprocessing resulted in one unsmoothed
$z$-statistic image for each fMRI resolution that summarized
task-related activation over the course of the functional scans.
We fit our model to the patient 1 $z$-statistic image data to compare
relative performance against a set of similar single-resolution
alternative methods. With this analysis, our goal was to show how our
method can be applied to identify peritumoral activations in
patient data and to illustrate potential benefits to inference using
combined spatial resolutions. In addition, we fit our model to the
$z$-statistic images from patient 2 to illustrate the method's
capacity to recover an estimate of activation in regions with signal
loss. Signal loss in fMRI data can occur where tissue types with
different magnetic field susceptibilities neighbor one another. This
is a common problem encountered in presurgical applications, and can
potentially lead to exclusion of areas of interest from the analysis
\cite[e.g.][]{haller2009pitfalls, stippich2015clinical}.

\subsection{Covariance estimation}
\label{subsec:analysis:kernel-estimation}

\begin{figure}[!htb]
\centering
\includegraphics[width=0.65\textwidth]{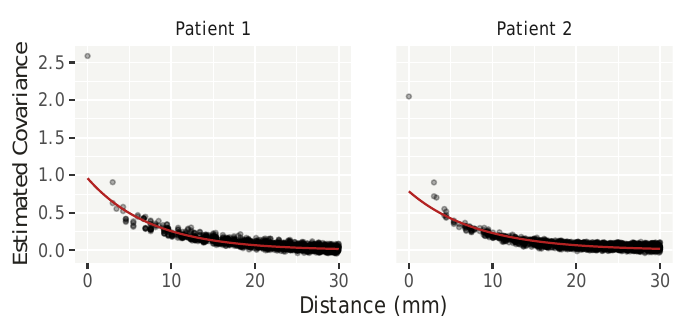}
\caption{Covariograms show empirical covariances between
  neighboring standard resolution voxels as a function of distance
  overlaid with a parametric estimate of the covariance function. 
\protect{\label{fig:covariogram}}}
\end{figure}

We chose to estimate the Gaussian process covariance hyperparameters
$\btheta$ in the spirit of empirical Bayes using the method of minimum
contrast. Minimum contrast estimation (MCE) originates from
\cite{diggle1981binary} as a moment estimation approach to spatial
modeling. The method seeks to estimate parameters of a function with a
known form by minimizing some discrepancy criterion given data. In our
case we extracted empirical covariances between voxels at different
distances (``empirical covariogram''). We then selected $\btheta$ to
minimize a nonlinear least squares objective over \eqref{eqn:rbf},
treating the empirical covariogram as pseudo data. Our Supplementary
Material \cite[][]{whiteman2022supp} gives a detailed overview of this
procedure for interested readers, as well as a brief sensitivity
analysis over our choice of covariance function in \eqref{eqn:rbf}.

This method is not without difficulty. For example, asymptotic theory
suggests that empirical covariogram estimation is
biased \cite[e.g.][]{cressie1984median}. Although this bias does not
decrease with increased sampling density (``infill asymptotics''), it
can be decreased by sampling data over increasing domains
\cite[][]{mardia1984maximum, stein1999interpolation,
  zhang2004inconsistent}. This point is worth acknowledging because
the kriging identity encoded in our prior
\eqref{eqn:approximate-prior} 
makes estimation of the correlation relatively important. On the other
hand, MCE is computationally efficient and scalable to large datasets,
and we found that it produced reasonable estimates of the true
covariance function in simulation (see the Supplementary Material).
One reason for this may be that with fMRI data we
have a tremendous amount of spatial information collected on a dense
grid. Although we can only estimate empirical covariances at a fixed
set of distances, we typically have tens of thousands of unique pairs
of voxels separated by those distances. In Fig. \ref{fig:covariogram},
we used the standard resolution images to estimate $\btheta$ as we
expect these data to have better SNR and there is no theoretical
benefit to adding infill locations as with the high resolution
images. In doing so, we make an appeal to the notion of a parent
process \cite[][]{banerjee2008gaussian} for $\pr\{ \mu(\cdot) \}$,
which could be defined such that in the prior
$\cov\{ \mu(\vox), \mu(\vox') \} = k(\vox, \vox')$ for all
$\vox, \vox' \in \Brain$.

For patient 1, an initial unrestricted estimate of $\btheta$ yielded
an estimated kernel exponent of $\nu \approx 1.25$; for improved
interpretability we reran our MCE procedure fixing $\nu = 1$ to yield
$\btheta = (0.887, 0.135, 1)\trans$. Optimization was performed using
the COBYLA algorithm \cite[][]{powell1994direct} as implemented by
\cite{nlopt} in the popular NLopt library. The resulting covariance
function is shown in the left panel of
Fig. \ref{fig:covariogram}, and  
corresponds to a 10.47 mm full width at half maximum (FWHM)
exponential correlation function. This estimate of $\btheta$ was used
for all of our analyses of patient 1's data; correspondingly, we set
the neighborhood radius $r$ to 10.35 mm in analyses of this patient's
data.

Similarly, we estimated $\btheta = (0.785, 0.132, 1)\trans$ for
patient 2. The resulting covariance function (also shown in
Fig. \ref{fig:covariogram}, right) corresponds to a 11.28 mm FWHM
exponential correlation function; we set $r$ to 11 mm for analysis of
this patient's data. In Fig. \ref{fig:covariogram}, the exponential
kernels appear to fit the empirical covariograms quite well. The
points at distances of 0 mm are not outliers but estimates of the
``sill'' or marginal variance of  the $\bY_s$ which in our model is
$\tau^2 + \sigma_s^2$. Consequentially, our algorithm constrains
$\tau^2$ to be strictly less than the empirical variance of $\bY_s$,
or whichever image is used to construct the covariogram.
In addition, the exponential models in
Fig. \ref{fig:covariogram} tend to mildly but systematically
underestimate the empirical covariances at displacements around 3
mm. We discuss how these data points can be modeled more accurately, 
and elaborate on why it may or may not be optimal to do so 
in the Supplementary Material \cite[][]{whiteman2022supp}.

\subsection{Patient 1: Inference on the functional signals}
\label{subsec:analysis:pat9-inference}

We fit our model to the data from patient 1 described in section
\ref{sec:data-methods} with custom software written in \textsf{C++}
that uses the Eigen \cite[][]{eigen} and FFTW \cite[][]{fftw}
libraries for linear algebra and DFT operations, respectively. For
these analysis, we set the number of leapfrog steps $L = 25$ and ran
three independent HMC chains of 4,000 iterations each, discarding the
first 1,000 as burnin, and thinning the output to every third iteration
thereafter. Univariate Gelman--Rubin statistics
\cite[][]{gelman1992inference} were used to evaluate voxelwise
convergence of $\bmu_h$. This statistic
was $\leq 1.03$ for every voxel, suggesting approximate
convergence. Additionally, trace plots of means from six randomly
selected voxels are shown in
in the Supplementary Material \cite[][]{whiteman2022supp}
and show good mixing of the Markov chains.

\begin{figure}[!htb]
\centering
\includegraphics[width=0.68\textwidth]{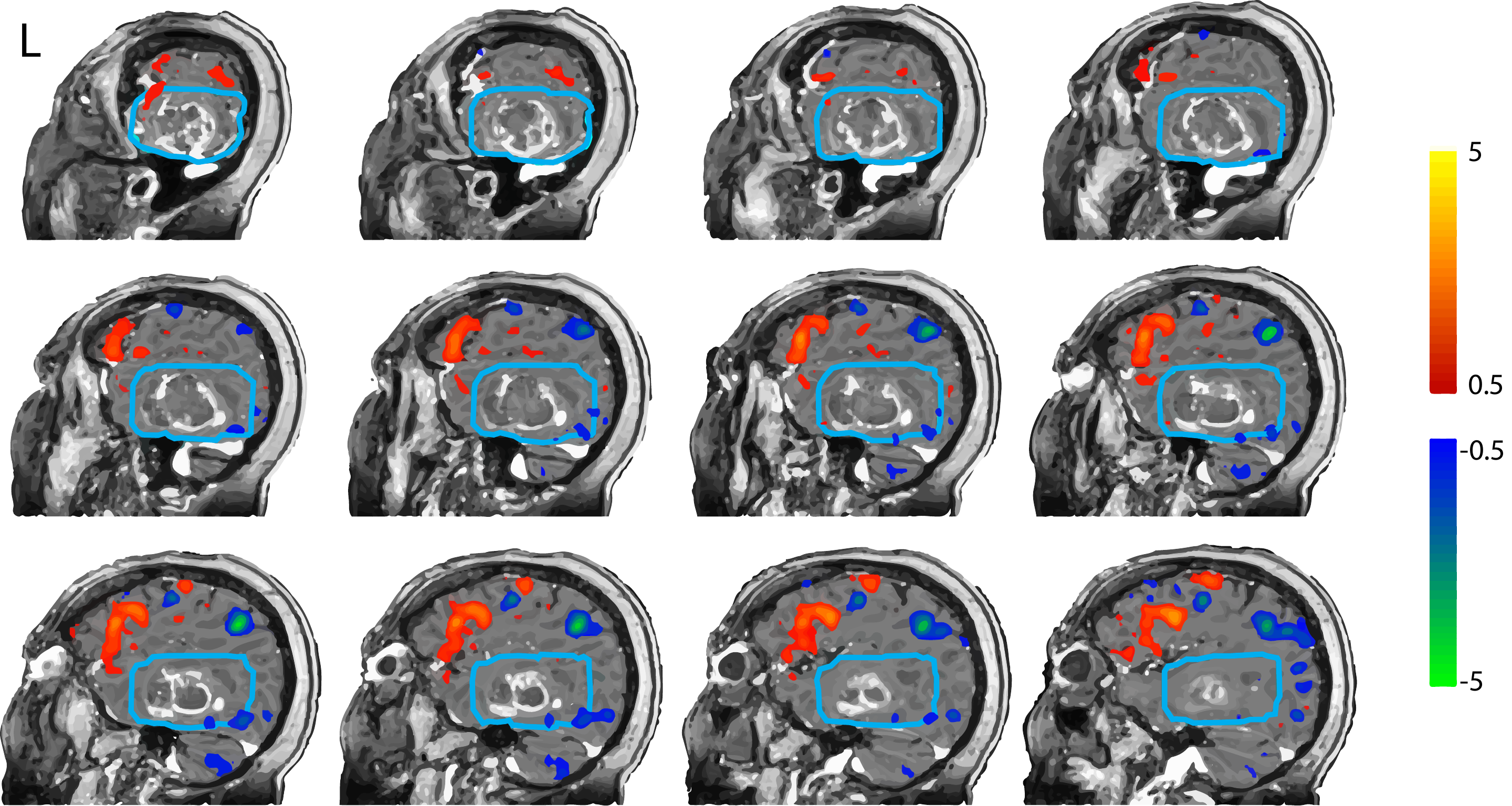}\hspace{0.02\textwidth}
\includegraphics[width=0.28\textwidth]{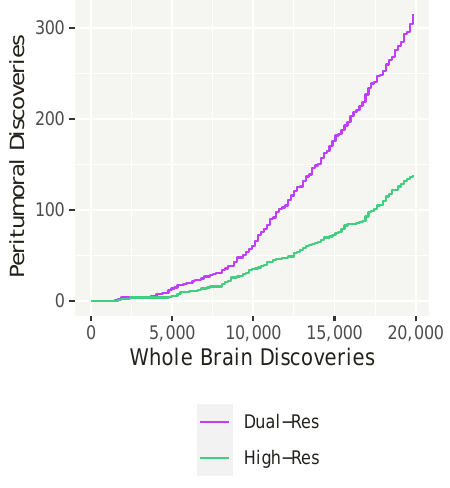}
\caption{Patient 1: (\emph{Left}) Thresholded posterior
  mean image shows peritumoral activation identified using our dual
  resolution method. The tumor is the region of mixed hypo- and
  hyperintensity in the temporal lobe across slices; the peritumoral
  region is outlined in each panel (in cyan). Functional activations
  are shown in warm colors, and functional deactivations are shown in
  cool colors, with units on the $z$-statistic scale. Activation
  regions are shown setting $k_1$ in our decision rule
  \eqref{eqn:decision-rule} 
  to 17 to enhance the visualization. Slices are shown
  proceeding lateral-to-medial through the left hemisphere in
  left-to-right, top-to-bottom order. (\emph{Right}) Cumulative counts
  of discoveries at varying decision thresholds. Voxelwise
  discoveries in the peritumoral region plotted against whole brain
  discoveries for both dual and high resolution methods.
  \protect{\label{fig:pat9-activation}}} 
\end{figure}

Fig. \ref{fig:pat9-activation} (\emph{left}) shows posterior mean
activation maps for a series of sagittal slices through the patient's
tumor in left temporal lobe. Activations are overlaid on a high
resolution, gadolinium enhanced T1-weighted anatomical scan.
In the figure, we have circled a peritumoral region that was deemed to
determine the surgical access considered. The patient's tumor can be
seen within this circled region in all slices.
As in our simulation studies, we compared performance of our dual 
resolution model against single resolution alternatives: models
considering only the high or standard resolution data, and an
additional setting using a naive average as data.

\begin{figure}[!htb]
  \centering
  \includegraphics[width=0.6\textwidth]{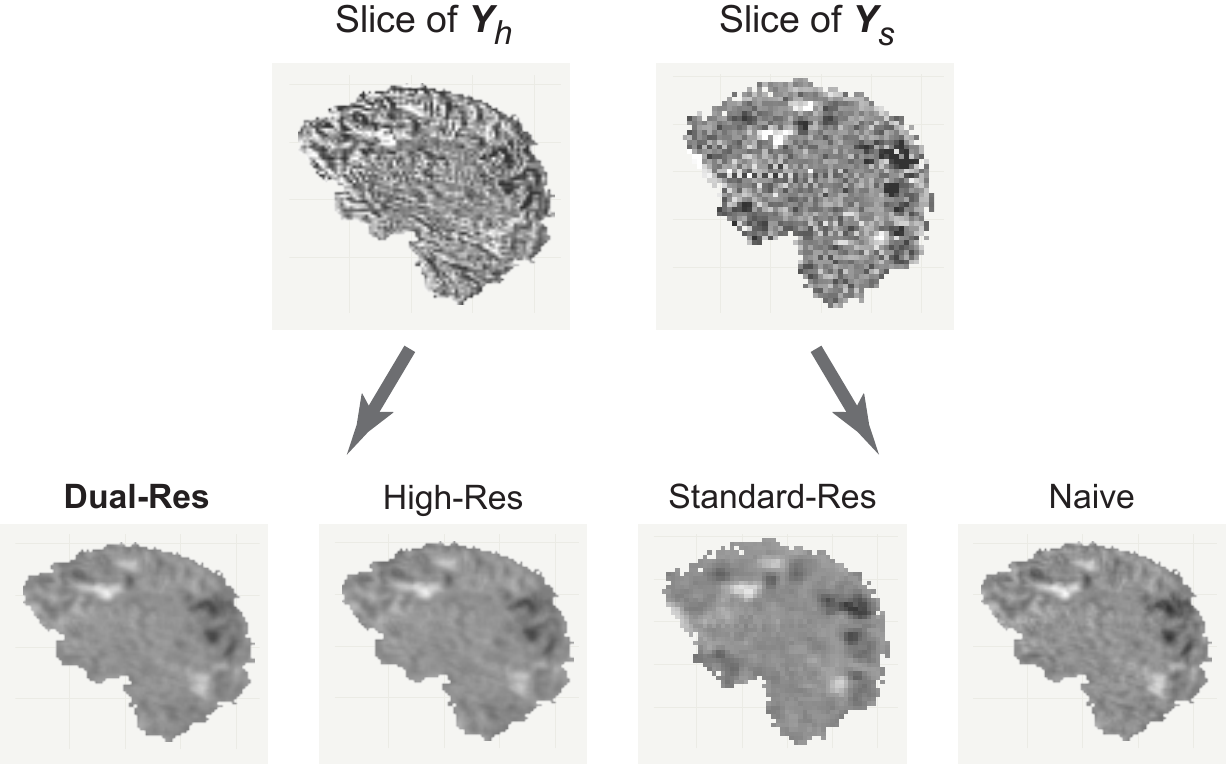}\vspace{1.5em}
  
  \includegraphics[width=0.31\textwidth]{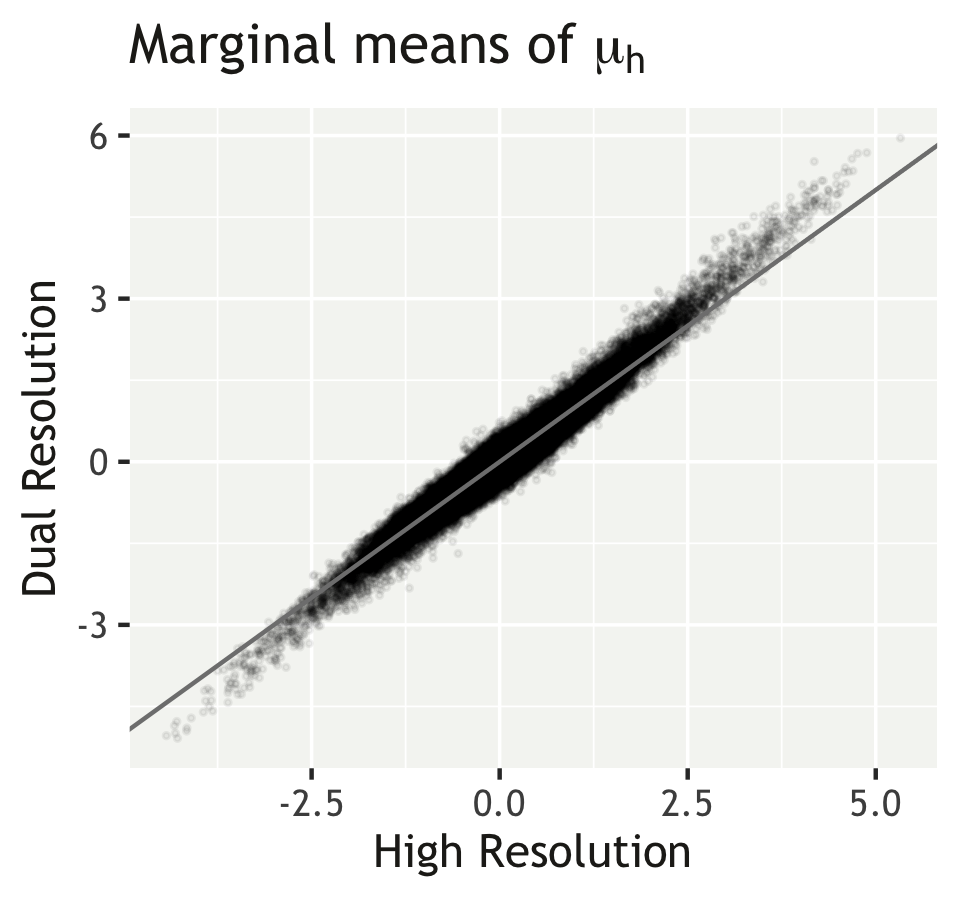}\hspace{2em}
  \includegraphics[width=0.31\textwidth]{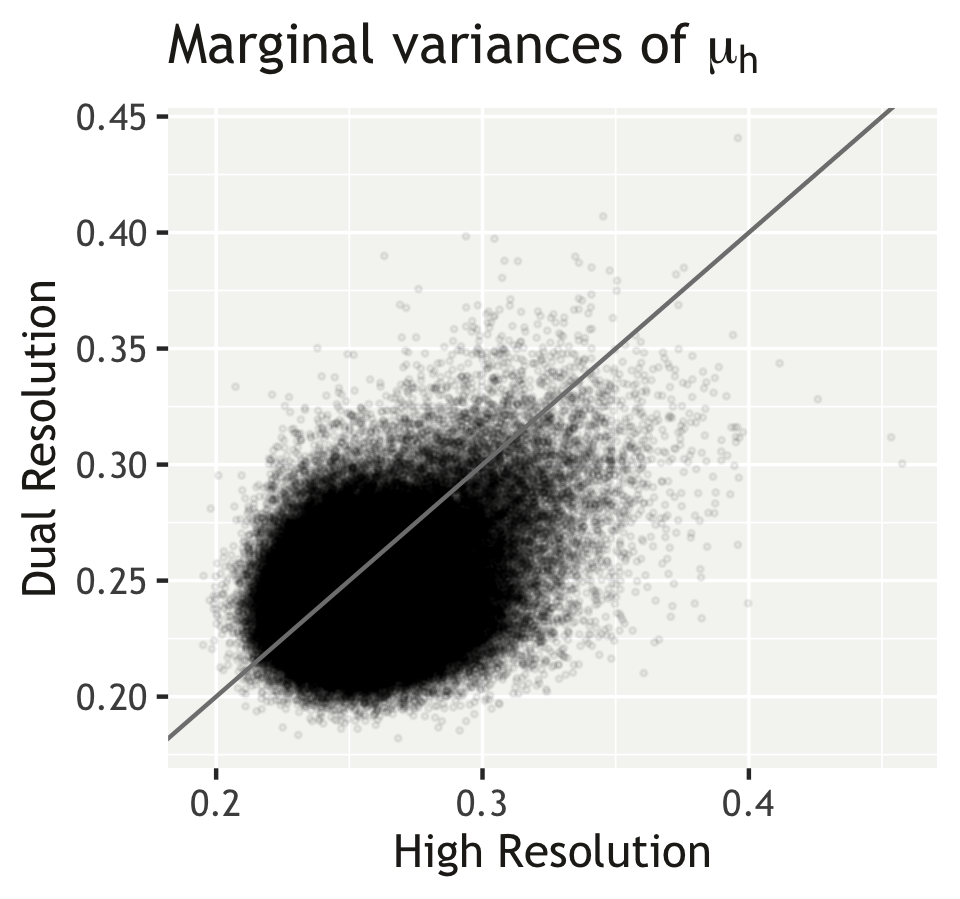}
  \caption{Patient 1: visual comparison of posterior means
    in a single sagittal slice from four models fit to different
    combinations of whole brain patient data (\emph{middle}). The
    (\emph{top}) row of the figure shows the raw data from the same
    slice at both high and standard resolution. Grayscale intensity is
    shared across all subfigures.
    The (\emph{bottom}) row shows a comparison of voxelwise posterior 
    means (\emph{bottom, left}) and variances (\emph{bottom, right})
    of the elements of $\bmu_h$ estimated using the proposed
    model and a single (high) resolution alternative. The gray lines
    show identity relationships for comparison; variances were lower
    using the dual resolution model in about 72.4\% of voxels.
    \protect{\label{fig:pat9-slice-comparison}}}
\end{figure}

In Fig. \ref{fig:pat9-activation} (\emph{right}), we show that no
matter the threshold applied to whole brain posterior activation maps,
our dual resolution method identified at least as many active voxels
in the peritumoral region than if we had ignored the standard
resolution data.
A visual comparison of the posterior mean of $\mu(\cdot)$ for
all four methods is shown in Fig. \ref{fig:pat9-slice-comparison}. We
chose a single sagittal slice to represent this comparison although
the analysis was over the whole brain. Qualitatively, posterior means
from the dual and high resolution analyses appear substantially
sharper than for the standard resolution analysis. At the same time,
differences are apparent in the dual and high resolution posterior
means, particularly around the edges of areas with high magnitude
signal. We also plot voxelwise comparisons of dual and high resolution 
posterior means and variances of $\bmu_h$ in
Fig. \ref{fig:pat9-slice-comparison}. 
In the figure, voxels with
high signal strength typically had higher magnitude posterior means
estimated with the dual resolution model; marginal variances of the
$\mu_{h,i}$, moreover, were lower with the dual resolution model
in about 72.4\% of voxels. With respect to mean image smoothness, we 
estimated (using our MCE procedure; see section
\ref{subsec:analysis:kernel-estimation}) the standard resolution
posterior mean image had a kernel 
FWHM of about 17.3 mm, and the high resolution posterior mean image had
a kernel FWHM of about 13 mm. Appropriately, the dual resolution
posterior mean image had a kernel FWHM between these two, at about 14.4
mm. Relating back to Fig. \ref{fig:dualres-problem}, our initial goal
in modeling joint data sources was to reduce noise inherent in the
high resolution signal and leverage signal strength from the standard
resolution data. Taken all together, these results demonstrate that we
have met that goal.

Additional patient 1 model fit and diagnostic evaluations are given in
the Supplementary Material \cite[][]{whiteman2022supp}.
In particular, we evaluated
the residual independence approximation present in our model likelihood
by running our kernel estimation procedure (see section
\ref{subsec:analysis:kernel-estimation}) on the model residual
images.
These analyses suggested that
residual correlation decayed to near zero within the smallest voxel
dimension widths, leading us to conclude that residual independence
was a reasonable approximation in our data. Full results are available
in the Supplementary Material \cite[][]{whiteman2022supp}.

\subsection{Patient 2: Recovery of lost signal}
\label{subsec:analysis:pat2-prediction}

\begin{figure}[!htb]
  \centering
  \includegraphics[width=0.2\textwidth]{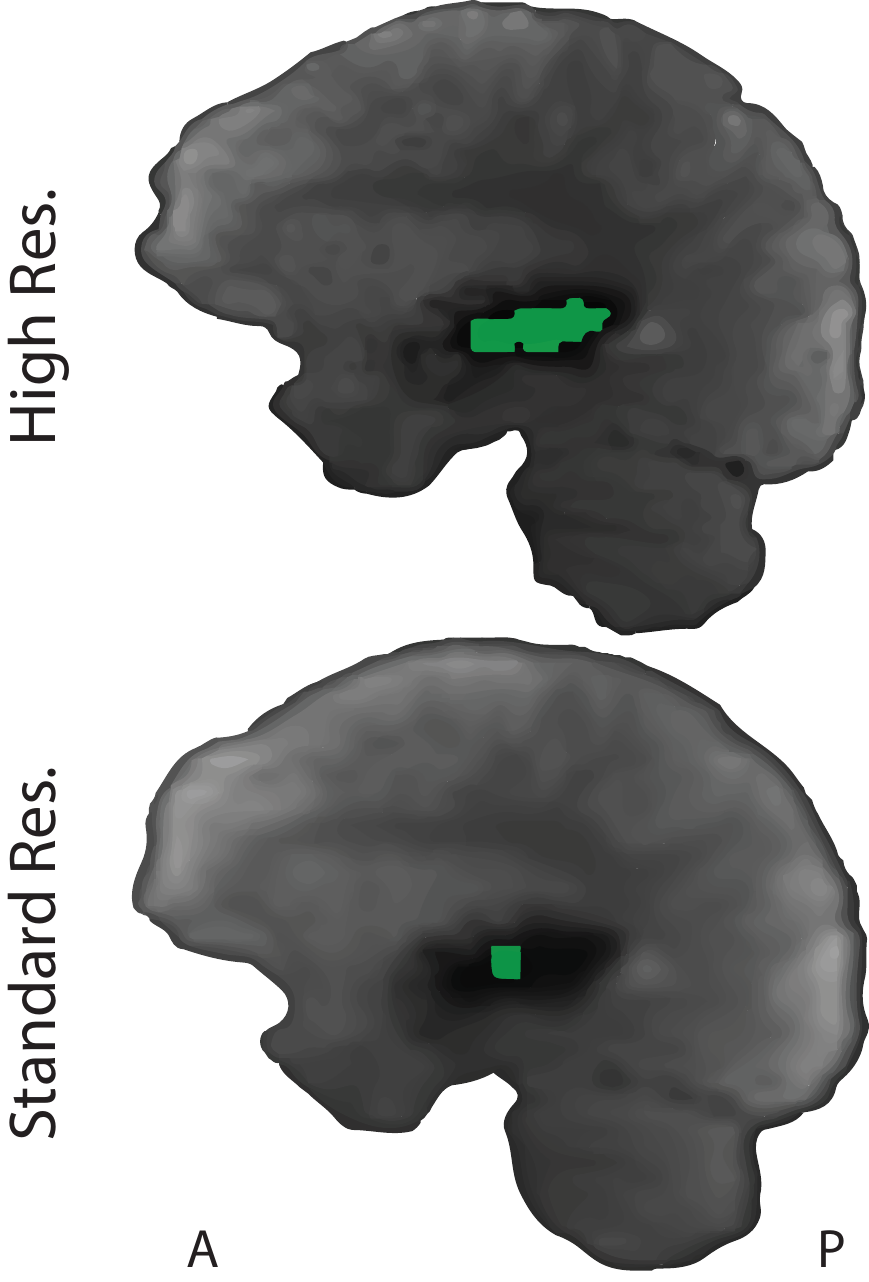}\hspace{0.05\textwidth}
  \includegraphics[width=0.45\textwidth]{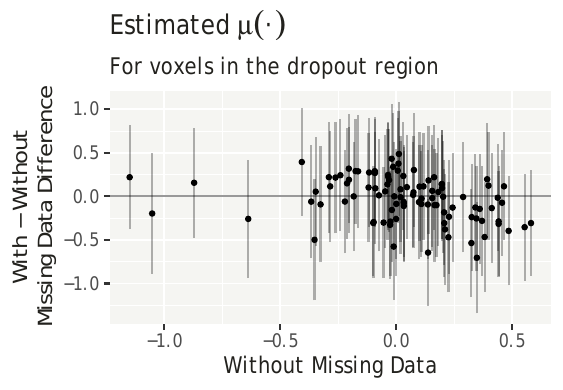}
  \caption{Patient 2: (\emph{Left})
    Core regions of fMRI signal loss across the left
    temporal and insular cortex are highlighted on 
    high and standard resolution T2$^*$-weighted slices.
    (\emph{Right}) 
    Comparison of the mean parameter for voxels in the core high
    resolution dropout region. We fit our dual resolution model
    to parallel versions of the data with and without missingness.
    The posterior mean estimate of $\mu(\cdot)$ without missing data
    is shown on $x$-axis, with the difference in the estimates
    shown on $y$-axis. Error bars give $\pm$ one standard error of the
    difference estimated across five HMC chains.
    \protect{\label{fig:holey-comparison}}}
\end{figure}

Similar to our analysis of patient 1, we fit our dual and single
resolution models to the data from patient 2. In this case, we ran
five independent HMC chains for each model and set the chain length,
burnin, thinning rate, and number of leapfrog steps identically as
above. Univariate Gelman--Rubin statistics were $\leq$ 1.05, again
suggesting approximate voxelwise convergence of $\bmu_h$. 
Our primary goal with this analysis was to illustrate our method's
ability to recover estimates of activation from regions of signal
loss.

In this particular data set, the patient's cavernoma caused a
region of GE-EPI signal dropout, with blooming along the left insular
and upper temporal lobe
(see Fig. \ref{fig:holey-comparison}, \emph{left}).
This is a common occurrence in clinical fMRI: brain lesions can induce
signal loss and magnetic susceptibility artifacts in gradient echo
imaging. The functional signal is
not ``missing,'' per se, but local signal hypointensities
can cause image preprocessing software to
exclude affected areas from analysis
\cite[][]{haller2009pitfalls}. Such was the case here, and we
leveraged this data structure to highlight our model's predictive
ability. By adjusting
brain/background thresholding in FSL, we created two versions of this
patient's task contrast data: one where data in the core dropout
region was completely masked out of all analyses (with missing data),
and one where voxels in this same region were included in all analyses
(without missing data). In this case, we were able to create parallel 
versions of the data for this patient, though in more
general practice it may sometimes be difficult to engineer the data
without missingness \cite[][]{haller2009pitfalls}.

We fit our models to both versions of the data to compare resulting 
functional estimates. In the case of missing data, the Gaussian
process formulation of our model enables natural prediction of values
of $\mu(\cdot)$ for voxels with missing data.
In the high resolution image, the region masked out due to signal
loss encompassed exactly 100 voxels, had a maximum length
(measured anterior to posterior) of around 25--26 mm, and was on
average about 9.4 mm wide (lateral to medial) and 4.5 mm tall (dorsal
to ventral). In the low resolution image, masked dropout was limited
to only two voxels given default brain/background thresholding in FSL.
With typical preprocessing pipelines we would generally expect more
signal loss voxels to be excluded from high resolution images.
In Fig. \ref{fig:holey-comparison} (\emph{right}) we show, for our dual
resolution method, the correspondence of predicted/estimated voxelwise
means in the dropout region for the two data variants.
The figure shows excellent correspondence:
the Pearson correlation between the predictions and the estimates is
0.673, suggesting 
our method has a strong capacity for signal recovery in dropout
regions of this size. Similar predictions/estimations using only the
high resolution data also show good correspondence, but were made
with higher variance relative to our dual resolution approach in 81 of
the 100 dropout region voxels.


Similarly, across the whole brain, the marginal
variances of $\bmu_{h}$ for patient 2 were lower with our dual
resolution approach in about 62.4\% of voxels (compared to the
high resolution model). Moreover, we estimated that the dual
resolution posterior mean image smoothness had a kernel FWHM of about
7.5 mm, while the high and standard resolution posterior mean image
smoothness FWHMs were about 7.6 mm and 13.5 mm, respectively.
Overall, estimation and inference about $\bmu_h$ was more similar
between our dual and high resolution models here than for the previous
patient.
Altogether, results from our analysis of the patient 2 data again
suggested our inference benefited from combining information from
both spatial resolutions, though the benefit may be less pronounced
compared to in patient 1. Based on our single data source high and
standard resolution alternative methods, we estimated that for patient
2, the standard resolution data provided only a modest 5.4\%
improvement in SNR compared to the high resolution data. By contrast,
for patient 1, the standard resolution data provided approximately
a 139\% improvement in SNR, lending context to the above result.



\section{Discussion}
\label{sec:discussion}

Preoperative fMRI presents many interesting and unique statistical
challenges from an applied perspective. Presurgical planning requires
spatially precise localization of patient specific functional
neuroanatomy, but the current physical limitations of MR imaging
technology lead to reductions in the signal to noise ratio (SNR) with
increases in spatial resolution. This inherent limitation has led to
the hypothesis that collecting fMRI data at multiple spatial
resolutions may result in improved functional region detection; our
simulations in the present paper suggest that this may indeed be the
case. We have also shown how a simple decision rule can be applied by
practitioners to infer about functional regions given some desired
trade off between false positive and false negative errors. This is 
important because neuroradiologists and neurosurgeons may
be more concerned with false negative errors, which could
lead to resection of functionally relevant tissue in
practice. 

With our present work, we propose to base inferences about
functional regions on a joint model for images collected at each
spatial resolution. Modeling high dimensional correlated outcomes can
be quite challenging computationally, and the dataset presented the
additional burden of integrating two data sources with different
spatial support sets. We circumvented this problem with a Gaussian
parent process approximation using only the highest collected resolution
image's voxel locations as a primary support set, and embedding these
locations within a larger, toroidal space, leading to computational
gains. As a consequence, our Gaussian process model and related
algorithm has very natural extensions to cases with different numbers
of data sources. For example, we recognize that not every preoperative
plan will rely on collecting both high and standard resolution fMRI
data. Our model can easily accommodate the situation where only one
spatial resolution is collected by simply dropping unobserved
data from the joint outcome.

Just as easily, our model could accommodate data collected at
additional spatial resolutions with minimal added computational
cost. In fact, in a different setting, we imagine our method could be
used for image based meta-analysis to synthesize results from multiple
experimental studies. In such a setting, posterior credible
sets---Bayesian analogs of spatial confidence sets from
\cite{bowring2021confidence}---could be used to shift inferential focus
back to limiting a family wise false positive error rate. With the
recent proliferation of large, multi-center imaging collectives
\cite[e.g.][]{vanhorn2009multi}, we feel this may be a promising area
for further applied research.

One important limitation of our present model is that we
treat both the prior mean model and the errors within each image as
stationary processes.
In general, stationarity may not be a realistic assumption
for imaging data \cite[e.g.][]{woolrich2004fully}. 
In the case of our mean model, stationarity is only a limitation of
the prior: given the data the posterior may still
reflect a non-stationary process.
In our analysis of the residual images from patient
data we found that while there was some residual spatial
autocorrelation, this autocorrelation in general decayed to near
zero within one to two voxel widths (see the Supplementary Material
for figures \cite[][]{whiteman2022supp}). Thus we concluded that prior
model mean-field stationarity can lead to reasonable posterior
approximations for these data. Our model on the error structure,
however, is a bit more restrictive.
We further note that residuals tended to show modestly higher
dispersion in gray matter than in white (see the Supplementary
Material \cite[][]{whiteman2022supp}). While we do not believe this
difference is so pronounced as to negatively impact our analyses, it
may be worthwhile to explore non-stationary error models.
In general, this is a difficult issue. Allowing too much flexibility
in the error process may, for example, lead to model
non-identifiability or similar complications.

In our simulations and analysis of patient data we showed that
our dual resolution method borrows strength from both data sources to
improve inference, especially around the edges of 
active regions, without sacrificing the spatial resolution of
the high resolution data (confer from
Figs. \ref{fig:pat9-slice-comparison} and \ref{fig:2d-inference}). To
accomplish this task with patient images, we started from the
output of typical single subject fMRI analyses, treating summary
statistics from voxelwise marginal time series models as
data. \cite{bowman2008bayesian} similarly used summary statistics from
voxelwise marginal models as data in a group analysis in an
experimental setting. Although their approach and setting was different
from ours, the authors also chose not to smooth their data during
preprocessing and made a similar independent noise approximation in
their model likelihood as we do here
\cite[][]{bowman2008bayesian}. We provide additional evaluation of our
independent and homogeneous residual noise approximations in the
Supplementary Material, 
and conclude that the approximations
are reasonable in our patient data.
While we might eventually like to
incorporate available time series information into our model, doing so
would only add to computational complexity, and it is unclear to what
extent spatial inference would improve as a result. At present, a
handful of integrated spatiotemporal models have been developed for
fMRI studies, but nearly all of these are intended to be fit to single
slice data, not whole brain
\cite[e.g.][]{penny2005bayesian, groves2009combined,
  lindquist2010adaptive}. Only more recently have variational
approximations been leveraged to enable whole brain spatiotemporal
inference at a reasonable computational cost
\cite[][]{siden2017fast}. In its current form, our work uses summaries
of temporal data to enable whole brain inference at a very fine
spatial scale, but there may be room to incorporate richer temporal
information into our model as part of future study.

Finally, in our work we estimated the Gaussian process hyperparameters
$\btheta = (\tau^2, \psi, \nu)\trans$ from the data in the spirit of
empirical Bayes. We accomplished this goal by minimizing a least
squares contrast function over an empirical covariogram estimated from
the data. Other approaches to learning these parameters include
maximizing the data marginal likelihood
\cite[e.g.][]{mardia1984maximum}, and fully Bayesian estimation
\cite[e.g.][]{banerjee2008gaussian}. We chose our minimum contrast
estimation (MCE) type approach as it is generally more extensible to
the size of our dataset. Computing the marginal likelihood would
involve inversion of an $(n \times n)$ matrix where $n$ is the number
of voxels or spatial locations. Our posterior computation algorithm
specifically avoids even constructing such a matrix, which is
impossible to store on most computer systems (see section
\ref{subsec:methods:prior-approximation}). Fully Bayesian estimation
of $\btheta$ on the other hand is possible, though
still computationally demanding. The kernel bandwidth and exponent
parameters, $\psi$ and $\nu$, respectively, can be quite slow to update
with multiple data sources, and computation time is a concern in a
preoperative setting. In contrast, the partial sill variance
$\tau^2$ is straightforward to update in our framework, and an
abundance of spatial data make this parameter strongly
identifiable. We considered updating $\tau^2$ by default in our
algorithm, but found that it did not dramatically affect spatial
inference in our data and sometimes led to slower Markov chain
mixing. As a result, we decided to condition inference on fixed
$\btheta$ by default in our analyses and consider alternative
estimation methods a possibility for future extension.

Conditional on $\btheta$, our method enables spatially precise
inference on whole brain fMRI data collected at multiple spatial
resolutions. Despite the very high dimensional nature of our data, our
method is computationally efficient enough to be viable for
application in presurgical planning. In addition, we have shown
through simulation that inference drawn from a joint model using both 
available data sources can lead to substantial improvement over
inference with single resolution alternatives. We hope that this body
of work will benefit the presurgical fMRI community, and may find
extension in experimental fields by supporting image based meta
analysis and results synthesis.




\begin{appendix}

\section{Details of fMRI data collection and preprocessing}
\label{app:fmri}

FMRI data collection and methods have been described previously
\cite[][]{liu2016pre}.
Briefly, the patients were scanned using a 3
Tesla TrioTim scanner (TQ engine, 32 channel head coil; Siemens
Medical Solutions, Erlangen) using gradient-echo echo-planar imaging
(GE-EPI; 3000 ms repetition time; 30 ms echo time; 0.69 ms echo
spacing; GRAPPA acceleration factor 2). High resolution structural T1
weighted MPRAGE and T2 weighted FLAIR scans were also acquired to aid
intraoperative neuronavigation and fMRI data preprocessing. The high
and standard spatial resolution scans largely followed the same
protocols, except that multi-band acceleration was used to increase
the spatial resolution of high resolution acquisitions while keeping
the temporal resolution the same between protocols (160 volumes were
collected for each run).

FMRI time series preprocessing without spatial smoothing was 
performed prior to our analysis using FSL software
\cite[version 6.0.4;][]{jenkinson2012fsl} and the FEAT tool
\cite[version 6.00;][]{woolrich2001temporal}.
Standard resolution fMRI data were padded by 8 voxels in x and y
(resulting in a $72 \times 72 \times 48$ grid), and high resolution
data by 10 voxels in z (resulting in $120 \times 120 \times 72$ grid). 
Given standard resolution voxel sizes of $3 \times 3 \times 3.45$
mm$^3$ (patient 1) and $3 \times 3 \times 3.3$ mm$^3$ (patient 2), and
high resolution voxel sizes of 
$1.8 \times 1.8 \times 2.3$ mm$^3$ (patient 1) and $1.8 \times 1.8
\times 2.2$ mm$^3$ (patient 2) this padding
ensured that standard and high resolution data spanned the same
field-of-view (FoV) within subject prior to further processing. The
difference in effective resolution between the two patients resulted
only from different interslice gaps (15\% for patient 1 vs. 10\% for
patient 2; interslice gap was lowered for patient 2 because of a
smaller head size). Optimal within-subject alignment
of the two runs was then achieved by downsampling the volume used as
the target reference for motion correction in the high resolution run
and supplying this downsampled image as an alternative reference image
for motion correction of the standard resolution time series.
Per FSL default, we used the middle volume of the recorded frames (the
80\textsuperscript{th} of our 160 volume time series) as the target
reference. This volume was downsampled to the gridding 
of the standard resolution run using FSLeyes (part of FSL)
using nearest-neighbor interpolation and no additional
smoothing.


Data were temporally filtered using a 0.011 Hz
high pass filter to remove low frequency drifts, and marginal linear
models were fit to the time series data at each voxel to create
summary statistic maps of task-related activation. In this last step,
task related regressors were convolved with the canonical hemodynamic
response function; temporal derivatives of resulting functions were
also used as covariates of no interest. Preprocessing resulted in one
unsmoothed $z$-statistic image for each fMRI resolution that
summarized task-related activation over the course of the scans. We
went on to use the generated test statistic maps as outcome data in
our subsequent analysis, treating the contrast images as noisy
measures of true activation.


\section{Cavernomas and additional details about Patient 2}
\label{app:cavernomas}

Cavernomas are a specific type of arteriovenous malformation
without shunting.
They contain closely apposed, angiogenetically
immature blood vessels, typically with intralesional bleeding
residuals. Cavernomas can be treated via microsurgical removal
\cite[e.g.][]{bertalanffy2002cerebral}; 
if left untreated, they may lead to seizures or progressive
neurological deficits upon symptomatic micro- or
macrohemorrhages. Our patient 2 was found to have a cavernous
malformation (cavernoma) with chronic and subacute hemorrhage
(Zabramski type I) in her left temporal lobe close to the transverse
temporal gyrus and insular cortex.


\end{appendix}

\section*{Acknowledgements}
We gratefully acknowledge the indispensable expert technical and
collaboration support provided by the MR application and collaboration
management teams of Siemens Healthcare GmbH, which enabled us to
record multi-band (i.e. simultaneous-multi-slice) -accelerated
acquisitions of high-resolution fMRI data at otherwise identical
parameter settings like for the standard spatial resolution runs. 
Dr. Andreas J. Bartsch has additional joint appointments with the
Department of Neuroradiology at the University of Wuerzburg,
Wuerzburg, Germany; and with the FMRIB Centre Department of Clinical
Neurology at the University of Oxford, Oxford, United Kingdom.
This work was partially supported by NIH R01 DA048993
(Kang and Johnson).

\begin{supplement}
  \stitle{Supplement to ``Bayesian Inference for Brain Activity from
    Functional Magnetic Resonance Imaging Collected at Two Spatial
    Resolutions''}
  \sdescription{The online supplement to this article provides
    additional details regarding posterior computation, MCMC and model
    diagnostics, and comprehensive simulation results.}
\end{supplement}

\begin{supplement}
  \stitle{Source Code Companion to ``Bayesian Inference for Brain
    Activity from Functional Magnetic Resonance Imaging Collected at
    Two Spatial Resolutions''}
  \sdescription{\textsf{C++} programs for analyses presented in this
    manuscript. High and standard resolution fMRI contrast data for
    Patient 1 is packaged along with the source code. Software is
    maintained at \url{https://github.com/asw221/dualres}.}
\end{supplement}


\begin{supplement}
\stitle{Posterior computation}
\sdescription{In Section
    \ref{subsec:methods:posterior-computation}, 
  we outlined a posterior computation algorithm for our model that
  relies on embedding the covariance of $\bmu_h$ in a higher
  dimensional nested block-circulant matrix. We present the details of
  this algorithm here.}
\label{supp:sec:computation}
\end{supplement}

Broadly, our posterior computation algorithm has a Hamiltonian
Monte Carlo (HMC) -within-Gibbs sampling structure. Full conditional
updates are available for all of our model parameters, but it is
numerically challenging to evaluate or sample from the full
conditional distribution of $\bmu_h$.

In the main text, we discussed how we drew inspiration from the work
of \cite{wood1994simulation} to design an efficient HMC algorithm to
facilitate sampling of $\bmu_h$. We elaborate on that algorithm in
detail here. First, we embed $\bmu_h$ in a higher
dimensional random field $\bu$, which is constructed so that the prior
variance of $\bu$ is a nested block-circulant matrix $\bC$. The prior
variance of $\bmu_h$---$\bK_h$---is a principal submatrix of $\bC$
(see Fig. \ref{fig:circulant-embedding} in the main text for a
schematic picture). We never actually construct or store the full
matrix $\bC$: its base $\bc$ can be computed following Algorithm
\ref{alg:compute-circulant-base} below. With only the base $\bc$ in
memory, the complex eigenvalues of $\bC$ can be computed using
discrete Fourier transform (DFT) software:
\[ \blambda \gets \fft(\bc) / N, \]
where $N$ is the number of elements in $\bc$.

Then, let $\bxi = \bu + \bv i$ represent a complex Gaussian random
field with real part $\bu$, imaginary part $\bv$, and with the 
prior properties that $\bu \perp \bv$ and
$\var(\bu) \equiv \var(\bv) \equiv \bC$.
Writing out the prior in terms of $\bxi$,
\[ \bxi = \bu + \bv i, \qquad \bu \sim \Gaussian(\bm{0}, \bC), \quad
  \bv \sim \Gaussian(\bm{0}, \bC), \]
does not change our  model, moreover: the imaginary and non-brain
parts of $\bxi$ can simply be integrated away to recover our original
prior on $\bmu_h$.
Similarly, our plan will be to obtain posterior samples of $\bxi$
through HMC, and then simply discard extraneous elements to be left
with a posterior sample of $\bmu_h$.

\begin{figure}[!ht]
\centering
\includegraphics[width=0.35\textwidth]{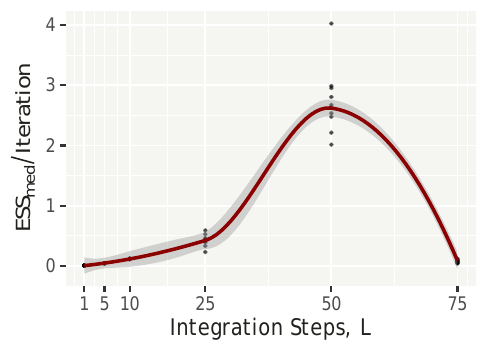}\hspace{0.05\textwidth}
\includegraphics[width=0.35\textwidth]{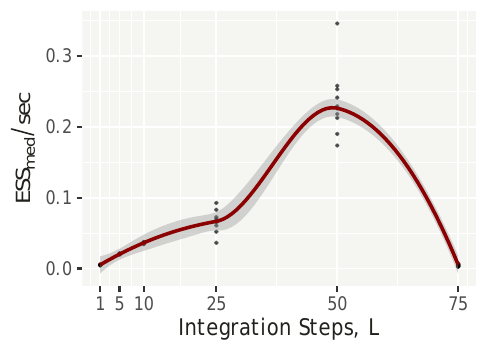}

  \begin{tabular}{ c c c }
    Method & Run time (hrs) & Total RAM (Gb) \\
    \hline
    Dual   & 2.76 & 2.32 \\
    High   & 1.88 & 2.05 \\
    Std    & 0.44 & 1.50 \\
    \hline
  \end{tabular}
\caption{Dual resolution algorithm efficiency (median ESS
  per iteration and per second) as a function of integration steps $L$
  in analysis of whole brain patient data.
  \emph{ESS} denotes the effective sample size of elements of
  $\bmu_h$. Peak efficiency was estimated around $L = 50$.
  Analyses were replicated 10 times for each value of $L$,
  and were timed on a Thelio System76 desktop with 62 Gb of free RAM
  and 20 logical cores (3.3 GHz Intel\textsuperscript{\textregistered}
  Core\textsuperscript{\textsf{TM}} i9 processors).
  Below the figure, we summarize the overall computational burden for
  real patient data on this hardware and at $L = 25$ steps.
  Run time is given in hours per 1,000 iterations; our naive method
  has the same cost as the high-resolution only method.
  \protect{\label{fig:leapfrog-efficiency}}} 
\end{figure}

HMC relies on several tuning parameters, including the choice of
momentum distribution, mass matrix, step size, and number of numerical
integration steps \cite[][]{neal2011hmc}. While a 
review of HMC-flavored algorithms and tuning parameter selection is
beyond the scope of this paper, we will detail our approach to tuning
parameter selection for model \eqref{eqn:gp-likelihood}. Given the other
tuning parameters and a target Metropolis-Hastings rate (which we
fixed at 65\%), we tuned the step size $\epsilon$ during warm up
following the dual averaging method of
\cite{hoffman2014nuts}. We then fixed $\epsilon_0$ at the
value of $\epsilon$ on the last burnin iteration, and drew $\epsilon
\sim \text{Uniform}(0.9\, \epsilon_0, 1.1\, \epsilon_0)$ to induce
random integration path lengths (the product $\epsilon L$) during
sampling, potentially helping the algorithm escape local modes
\cite[][]{neal2011hmc}. To inform selection of the number of leapfrog
integration steps $L$, we performed repeated analyses of patient data.
Results of this experiment
suggest $L = 25$ or $L = 50$ as practical starting points for best
algorithmic efficiency (see Fig. \ref{fig:leapfrog-efficiency}).

Let,
\[ \mathcal{L}(\bxi) = \ln \pr(\bxi \mid \bY_h, \bY_s, \bmu_s, \btheta, 
  \sigma_h^2, \sigma_s^2, r) \]
represent the full conditional log posterior of $\bxi$. Since
$\exp\{ \mathcal{L}(\bxi) \}$ is complex Gaussian, we in turn chose a
complex Gaussian distribution for HMC momenta.
\cite{girolami2011riemann} suggest exploiting Riemannian
geometry in HMC by adapting the algorithm's mass matrix, $\bM$, to the
local curvature of the log posterior. The authors suggest that taking
$\bM$ proportional to the negative Hessian of the log posterior leads
to improved algorithmic efficiency in high dimensions, though this
approach is not typically feasible when the dimension of $\bM$ is more
than a few thousand. Up to a permutation of $\bxi$, in our model, we
have that,
\begin{equation}\label{eqn:hessian-xi}
  -\nabla^2 \mathcal{L}(\bxi) =
  \begin{pmatrix} \sigma_h^{-2} \bI + \sigma_s^{-2} \bW\trans \bW
    \\ \bm{0}
  \end{pmatrix} + \bF \bLambda^{-1} \bF\adj,
\end{equation}
where $\bLambda = \diag(\blambda)$, and $\bF$ is the 3D DFT matrix as
in the main text. 
In the present case,
\eqref{eqn:hessian-xi} is ultrahigh dimensional and impossible to work
with directly, but by dropping the term involving $\bW\trans \bW$,
which is dense, and extending $\sigma_h^{-2} \bI$ we can arrive at an
alternative choice of mass matrix. Let $\bM(\sigma_h^2)$ denote the
matrix-valued function, 
\begin{equation}\label{eqn:mass-matrix}
\bM(\sigma_h^2) = \bF [\bLambda^{-1} + \sigma_h^{-2} \bI] \bF\adj,
\end{equation}
which, like $\bC$, is nested block-circulant, and easy to compute
with. Circulant matrices have been used successfully as
preconditioners in other gradient-based optimization schemes for
imaging problems \cite[e.g.][]{fessler1999conjugate}. 
If each element in $\Re(\blambda)$ is strictly
greater than zero, $\bM(\sigma_h^2)$ is positive definite and so can
be used to define a metric tensor on a Riemannian manifold as in
\cite{girolami2011riemann}. Some additional intuition can be gained by
considering how \eqref{eqn:hessian-xi} relates to a missing data
problem. If we were only modeling high resolution data, and those data
were observed on the entire extended grid with variance $\sigma_h^2$,
then \eqref{eqn:mass-matrix} would be exactly the negative Hessian of
$\mathcal{L}(\bxi)$.

\begin{algorithm}[H]
\caption{Riemann manifold HMC for dual resolution mapping
  models. $\bF$ denotes the scaled 3D DFT matrix; products of the
  form, $\bF\adj \bp = \ifft(\bp)$, for example, can be computed
  efficiently using DFT software. 
  \protect{\label{alg:hmc}}} 
\begin{algorithmic}[1]
%
\Procedure {UpdateMean}{$\bxi$; $\blambda$, $\sigma_h^2$, $\epsilon$,
  $L$} 
\State Compute eigenvalues of $\bM(\sigma_h^2)$: 
    \State\hspace{\algorithmicindent} $\lambda_i^M \gets \sigma_h^{-2}
    + \lambda_i^{-1}$ 
    \State\hspace{\algorithmicindent} Set $\bLambda_M \gets
    \text{diag}\{ \lambda_i^M \}$, $i = 1, \ldots, \text{dim}(\bxi)$ 
\State Sample momentum, $\bp \sim \mathcal{CN}(\bm{0}, \bF
\bLambda_M \bF\adj)$ 
\State Compute total energy, $H \gets -\mathcal{L}(\bxi) + \frac{1}{2}
\bp\adj \bF \bLambda_M^{-1} \bF\adj \bp$ 
\State Set $\bxi^{\text{new}} \gets \bxi$
\For{$l$ in $1, \ldots, L$}
    \Comment{Leapfrog integrator}
    \State $\bp \gets \bp + \frac{\epsilon}{2} \nabla
    \mathcal{L}(\bxi^{\text{new}})$ 
    \State $\bxi^{\text{new}} \gets \bxi^{\text{new}} + \epsilon \bF
    \bLambda_M^{-1} \bF\adj \bp$ 
    \State $\bp \gets \bp + \frac{\epsilon}{2} \nabla
    \mathcal{L}(\bxi^{\text{new}})$ 
\EndFor
\State Compute $H^{\text{new}} \gets -\mathcal{L}(\bxi^{\text{new}}) +
\frac{1}{2} \bp\adj \bF \bLambda_M^{-1} \bF\adj \bp$ 
\State Set $\bxi \gets \bxi^{\text{new}}$ with probability $\alpha =
\min \{ 1, \exp(H - H^{\text{new}}) \}$
\State Discard all elements of $\bxi$ that do not correspond to
$\bmu_h$ 
\State Return posterior sample of $\bmu_h$
\EndProcedure
\end{algorithmic}
\end{algorithm}

With all this in hand, samples of $\bmu_h$ can be drawn following 
Algorithm \ref{alg:hmc}. In particular, note how all products
involving $\bF$ can be computed with DFT software. In addition, the
quadratic forms in Algorithm \ref{alg:hmc} represent computations over
ultrahigh dimensional components. The quadratic form
$\bp\adj \bF \bLambda_M^{-1} \bF\adj \bp$, for example, can be
evaluated by computing,
\[ \bphi \gets \ifft(\bp), \]
into a temporary product, $\bphi$, and then summing over terms
$\sum_i \bar{\phi}_i \cdot \phi_i / \lambda^M_i$, where $\bar{a}$
denotes the complex conjugate of $a$.
When working in single precision, we
found it necessary to use the Kahan summation algorithm
\cite[][]{kahan1965pracniques}, or similar correction, to evaluate
these long sums accurately.

Finally, our other parameters, $\bmu_s$, $\sigma_h^2$, and
$\sigma_s^2$ can easily be sampled with full conditional Gibbs
updates. 
We particularly note that our prior places the restriction
$\sigma_h^2 > \sigma_s^2$
so that the conditional posteriors of both nugget variance parameters
are truncated inverse Gamma. We sometimes encountered numerical
difficulty sampling these parameters during warm up. As a result, we
chose to ignore the restriction on $\sigma_h^2$ and $\sigma_s^2$
programmatically, and simply discard posterior samples where the
restriction was not satisfied. After warm up, however, we found that
even when working with patient data the posterior probability
that $\sigma_h^2 > \sigma_s^2$ was effectively unity, and that we
never had to discard or post-process MCMC samples in this way.


\begin{supplement}
\stitle{Circulant base construction}
\sdescription{This section presents a simple algorithm to illustrate
  circulant matrix base computation for our applications.}
\label{supp:sec:circulant-base}
\end{supplement}

\begin{algorithm}[H]
\caption{Compute the base of a circulant matrix associated with a 3D
  grid 
  \protect{\label{alg:compute-circulant-base}}} 
\begin{algorithmic}[1]
\Procedure {ComputeCirculantBase}{$\bd$, $K(\cdot, \cdot;
  \bm{\theta})$} 
\State Inputs: $\bd$, original 3D grid dimensions; $K(\cdot, \cdot;
\bm{\theta})$ covariance function parameterized by $\bm{\theta}$ 
\State Compute extended grid dimensions, $d^\star_i \gets 2^{\log_2
  \lceil 2 (d_i - 1) \rceil}$ for $i = 1, 2, 3$ 
\State $k \gets 0$, $h \gets 1$ 
\State Find location $\vox_1$ associated with grid position $(1, 1,
1)$ 
\For{$l$ in $1, \ldots, d^\star_3$} 
\Comment{Column-major order}
    \State $j \gets 0$
    \State \textbf{if} {$l \leq d_3$} \textbf{then} $k \gets k + 1$
    \textbf{else} $k \gets k - 1$
    \For{$m$ in $1, \ldots, d^\star_2$}
        \State $i \gets 0$
        \State \textbf{if} {$m \leq d_2$} \textbf{then} $j \gets j + 
        1$ \textbf{else} $j \gets j - 1$
        \For{$n$ in $1, \ldots, d^\star_1$}
            \State \textbf{if} {$n \leq d_1$} \textbf{then} $i \gets i 
            + 1$ \textbf{else} $i \gets i - 1$
            \State Find location $\vox$ associated with grid position
            $(i, j, k)$ 
            \State Compute $c_h \gets K(\vox_1, \vox; \bm{\theta})$ 
            \State $h \gets h + 1$
        \EndFor
    \EndFor
\EndFor
\State Return circulant matrix base, $\bc$
\EndProcedure
\end{algorithmic}
\end{algorithm}

\begin{supplement}
  \stitle{Analysis of patient 1 data: Model diagnostics}
  \sdescription{
    In this section we include several of the general attempts we have
    made to probe Markov chain convergence and model fit in our
    analysis of patient data.
  }
\label{supp:sec:pat9-diagnostics}
\end{supplement}

\begin{figure}[!htb]
\centering
\includegraphics[width=0.75\textwidth]{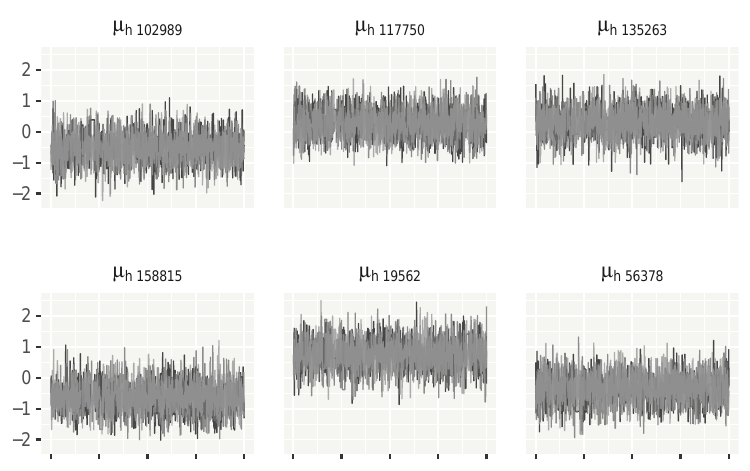}
\caption{Trace plots for the mean parameter of six random voxels from
  analysis of patient 1's data with our dual resolution model. Three
  different HMC chains are overlaid on one another in each subfigure.
  \protect{\label{supp:fig:pat9dual-traceplots}}}
\end{figure}

\begin{figure}[!htb]
    \centering
    \begin{tabular}{ c c }
        \textsf{Dual-Res} & \textsf{High-Res} \\
        \includegraphics[width=0.25\textwidth]{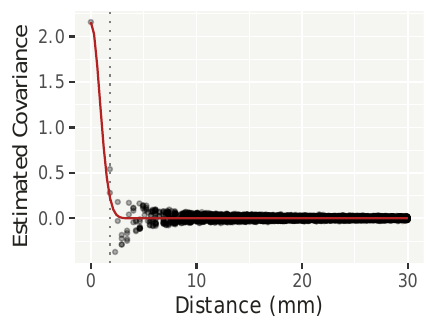}
                          & 
        \includegraphics[width=0.25\textwidth]{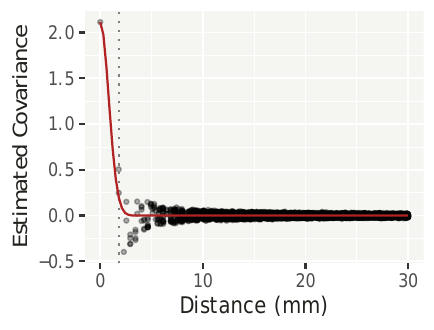}
      \\ 
        \textsf{Naive} & \textsf{Std-Res} \\
        \includegraphics[width=0.25\textwidth]{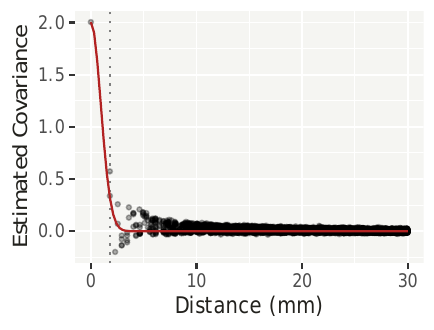}
                          & 
        \includegraphics[width=0.25\textwidth]{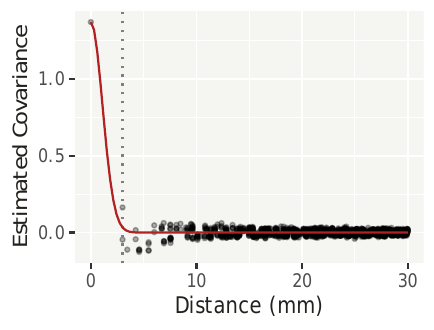}
    \end{tabular}
    \raisebox{-0.5\height}{\includegraphics[width=0.32\textwidth]{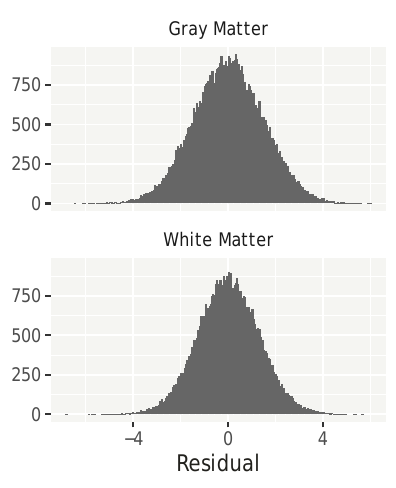}}
    \caption{(\emph{Left}) Residual covariograms for each method. The
      dotted lines show minimum voxel dimensions for each resolution,
      suggesting that the residual independence approximation is
      reasonable in these data. (\emph{Right}) Dual resolution method
      residual histograms roughly separated by gross tissue
      type. Residuals have modestly higher dispersion in gray
      matter than in white. 
      \protect{\label{supp:fig:residual-covariograms}}}
\end{figure}

As discussed in the main text,
Fig. \ref{supp:fig:pat9dual-traceplots} shows trace plots for three
chains of Hamiltonian Monte Carlo (HMC) draws of the mean parameter
for six random voxels. In all cases we examined, chains appear to show
good convergence and mixing. Fig. \ref{supp:fig:residual-covariograms}
(\emph{left}) shows empirical covariograms and estimated covariance
functions for residual images from each method. We found that the 
estimated residual correlation functions' full widths at half maxima
were on the order of the minimum voxel dimensions in all cases.
These analyses suggested that residual
correlation decayed to near zero within the smallest voxel dimension
widths, leading us to conclude that residual independence was a
reasonable approximation in our data.

The right panel of Fig. \ref{supp:fig:residual-covariograms} shows
histograms of the residuals from our dual resolution method roughly
separated by gross tissue type. We chose to parse the residuals in
this way due to some concern that a homogeneous residual variance
approximation may not be fully justifiable across the whole brain. 
To construct this figure, we created non-overlapping gray and white
matter tissue labels using the FAST program from the FSL software
suite \cite[][]{fast}, though the presence of the tumor complicates
this procedure. The figure suggests that residuals had modestly higher
dispersion in gray matter (standard deviation = 1.49) than in white
(standard deviation = 1.31). If it were not for the tumor, we might
ideally only want to analyze gray matter voxels for signs of
task-related activation. Given the present context, however, this
strategy is not completely possible. As it stands, although it appears
homogeneous residual variance may not strictly hold across different
tissue types, we do not believe the approximation is so poor as to
grossly impact our analyses in a negative way.

We further examined posterior predictive distributions for the data
from each voxel in the high resolution image, and compared the
distributions against the observed data (analysis not shown). Dual
resolution model posterior predictive inverse quantiles for the
observed data were roughly uniform, suggesting that data outliers
occurred no more or less frequently than would be expected given the
model.

\begin{supplement}
  \stitle{Analysis of patient 2 data: Sensitivity analysis}
  \sdescription{In this section we include a brief sensitivity
    analyses related to the choice of neighborhood size
    and covariance function in our dual resolution method.}
  \label{supp:sec:pat2-sensitivity}
\end{supplement}

\begin{figure}[!htb]
  \centering
  \includegraphics[width=0.3\textwidth]{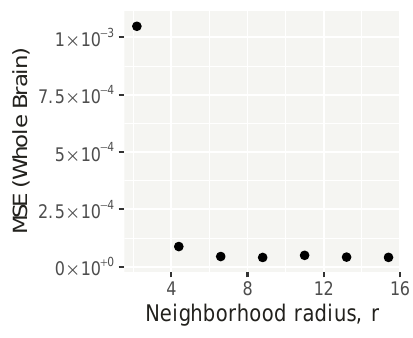}\hspace{0.05\textwidth}
  \includegraphics[width=0.3\textwidth]{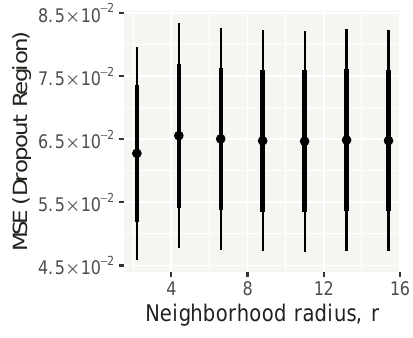}
  \caption{Mean squared error (MSE) of the posterior
    expectation of $\mu(\cdot)$ given fixed $\btheta$ but different
    values of $r$. The (\emph{left}) panel shows MSE of $\mu(\cdot)$
    evaluated across the whole brain, while the (\emph{right}) panel
    shows the predictive MSE for voxels in patient 2's dropout
    region. Thick and thin lines give approximate 80\% and 95\%
    confidence intervals.
    \protect{\label{supp:fig:pat2-neighborhood-sensitivity}}}
\end{figure}

Our dual resolution mapping method relies on a neighborhood radius
parameter $r$ to construct locally kriged samples of $\bmu_s$ given
$\bmu_h$ (see section \ref{subsec:methods:prior-approximation} in the
main text). Conceptually, this construction is somewhat inspired by
the so called nearest neighbor Gaussian process 
\cite[][]{datta2016hierarchical,finley2019efficient}.
In practice, we treat $r$ as a hyperparameter and condition
analyses on it, though it is of interest to understand how the choice
of $r$ affects inference about $\bmu_h$.
In our patient data analyses (sections
\ref{subsec:analysis:pat9-inference} and
\ref{subsec:analysis:pat2-prediction} in the main text), we set $r$ to
be approximately equal to the estimated full kernel width at half
maximum. This resulted in neighborhood sizes on the order of 300--700
voxels for our patient data.

To explore the influence of $r$ on estimation and prediction, we fit
our dual resolution model to the patient 2 data under several
different settings, all for fixed $\btheta$. As a comparison point, we
took the posterior mean of $\mu(\cdot)$ fit to the data without
missingness and conditioned on $r = 11$ mm. We then compared against the
posterior mean of $\mu(\cdot)$ from repeat analyses of the
with-missingness data and varying values of $r$ (see section
\ref{subsec:analysis:pat2-prediction} in the main text for an
explanation of the two data sets). For these repeat analyses, we chose
values of $r$ based on multiples of the largest high resolution image
voxel dimension (2.2 mm).
Fig. \ref{supp:fig:pat2-neighborhood-sensitivity} summarizes this
experiment in terms of the squared error of $\mu(\cdot)$ averaged over
the whole brain (\emph{left}) and voxels in the dropout region
(\emph{right}). From these results we conclude that as long as $r$ is
sufficiently large ($\geq$ 6.6 mm or so; corresponding to
neighborhood sizes of at least 100--200 voxels), it does not appear to
have much influence on posterior estimates.

A reviewer pointed out that, especially in our analysis of Patient 2's
covariogram, the exponential model we used tends to underestimate the
proximal empirical covariances. 
We chose to use the radial basis covariance function-family largely
because of the substantial history of gaussian smoothing in applied MRI
analysis. Additional literature suggests exponential smoothing kernels
are perhaps more appropriate for fMRI data
\cite[][]{groves2009combined}.
The reviewer's note prompted us to consider alternative covariance
functions and their impact on our analysis, and we summarize one
alternative here.

\begin{figure}[!htb]
  \centering
  \includegraphics[width=0.35\textwidth]{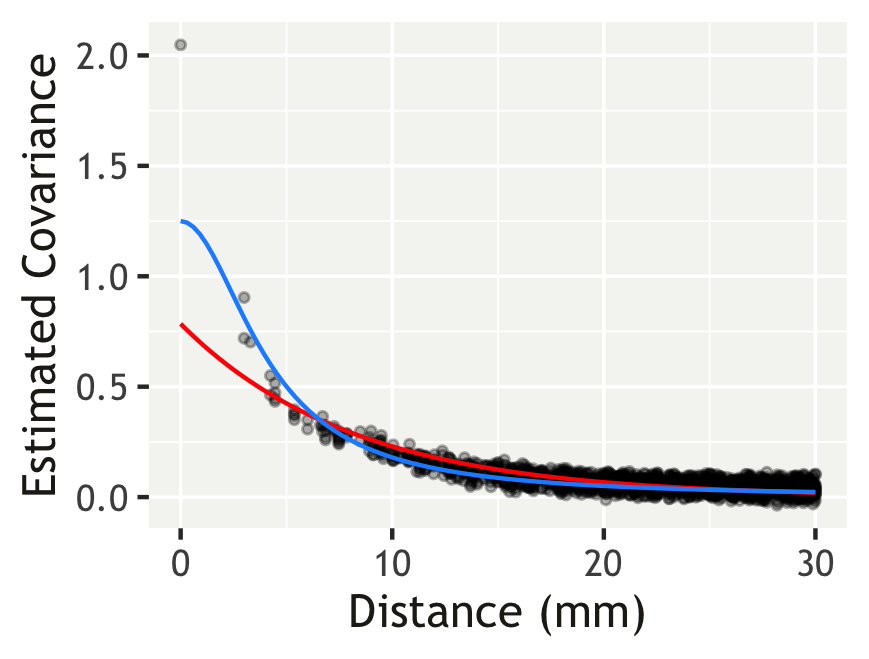}
  \caption{Reanalysis of Patient 2's covariogram. The red line
    reproduces the exponential covariance model from the main text;
    the blue line shows a rational quadratic covariance model.
    \protect{\label{fig:pat2-mce-reanalysis}}}
\end{figure}

In Fig. \ref{fig:pat2-mce-reanalysis}, we compare the exponential
covariance model from the main text against a rational quadratic
model, and find that the rational quadratic model fits the proximal
empirical covariances quite well. The specific rational
quadratic model in the above figure is,
\[ k_{R.Q.}(\vox, \vox') = 1.25 \bigg( 1 + \frac{\lVert \vox -
    \vox'\rVert^2}{16.67 \times 0.99} \bigg)^{-0.99}. \]
It is impossible to tell visually, but the rational quadratic model in
Fig. \ref{fig:pat2-mce-reanalysis} is \emph{sub}-optimal in the sense
that it has a very slightly higher residual weighted sum of squares
than the exponential model. Better than either might be some weighted
linear or piecewise combination of the two.

Choice of the covariance function is more art than science.
An interesting feature of this problem is that the empirical
covariances will tend to overestimate the true mean field covariance
if the noise is positively correlated spatially. 
Assuming as we do in the main text that
$Y(\vox) = \mu(\vox) + \epsilon(\vox)$ and that
$\mu(\vox) \perp \epsilon(\vox')$ for all $\vox, \vox'$, the empirical
covariances will be overestimates since,
\[ \cov\{Y(\vox), Y(\vox')\} = \cov\{\mu(\vox), \mu(\vox')\} +
  \cov\{\epsilon(\vox), \epsilon(\vox')\}. \]
Although in our work we modeled the error structure as
a white noise-type process for simplicity, it is perhaps more
realistic to assume the errors may be positively correlated over short 
distances. If the errors are in fact correlated spatially, then it may
be preferable to use a covariance model that underestimates the
proximal empirical covariances.

\begin{figure}
  \begin{tabular}{ c c }
    Exponential & Rational Quadratic \\
  \includegraphics[width=0.2\textwidth]{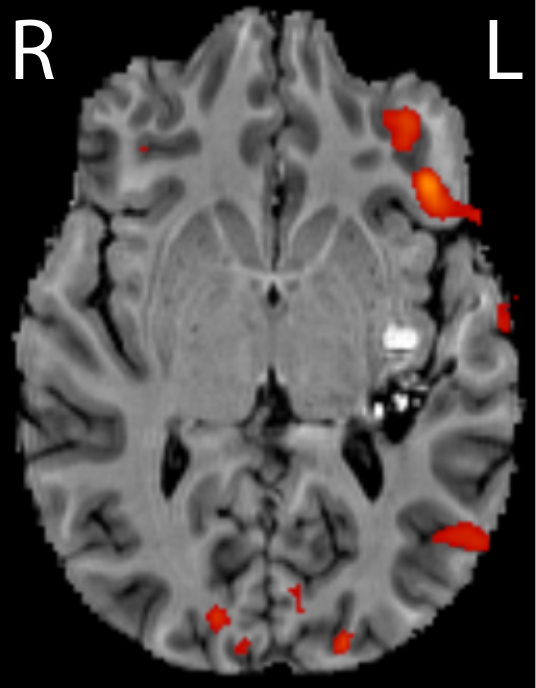} &
  \includegraphics[width=0.2\textwidth]{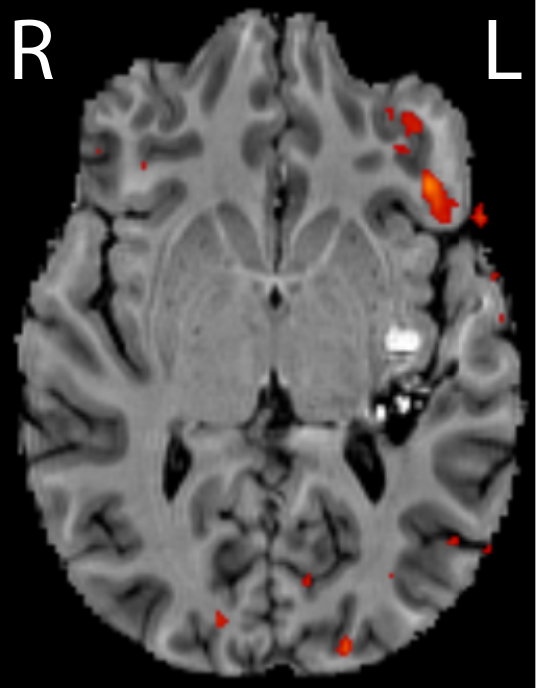} 
  \end{tabular}
  \caption{The figure shows thresholded posterior inference of
    activation regions for Patient 2 in an example horizontal
    slice. The color scale is shared between sub figures and reflects
    an approximate posterior probability of activation (range
    0.3--1.0).
    \protect{\label{fig:pat2dual-reanalysis-comparison}}}
\end{figure}

In practice we observed mild, intermittent spatial autocorrelation
patterns in our residual images (see Section
\ref{supp:sec:pat9-diagnostics} for example).
We switched from the exponential to the rational quadratic
covariance function and did not find that this change ameliorated 
residual autocorrelation patterns. Because the rational quadratic
covariance model places higher correlation between proximal elements
and decays more rapidly, we found that it yielded posterior inference
that was both noisier and less sensitive than our primary analysis
(see Fig. \ref{fig:pat2dual-reanalysis-comparison}).
Clearly results are sensitive to the choice of covariance function to
some degree, underscoring the importance of these issues.

\begin{supplement}
  \stitle{2D simulation results}
  \sdescription{
    In this section we give results designed augment those reported in
    section \ref{subsec:simulations:2d-recovery} with additional 
    simulation settings. Tables follow the exact format of Table 
    \ref{tab:2d-sims-main} in the main text.
  }
\label{supp:sec:2d-sims}
\end{supplement}

In all cases
considered, our dual resolution method had the lowest mean squared
error (MSE) and false negative rate. Of potential interest, however,
is that the high resolution-only method was the second best performer
when data were simulated with a marginal exponential correlation
structure (Table \ref{supp:tab:2d-sims-exponential}), and the naive
data averaging method was the second best performer when data were
simulated with a marginal Gaussian correlation structure (Table
\ref{supp:tab:2d-sims-gaussian}).

\begin{table}[ht]
\centering
\begin{tabular}{ l l c c c l }
  \multicolumn{1}{c}{\emph{Model}} 
  &  \multicolumn{1}{c}{\emph{Kernel}} 
  & \emph{SNR}$_s$\emph{:SNR}$_h$
  & \emph{SNR}$_h$
  & \emph{MSE}
  & \multicolumn{1}{c}{\emph{False --}} \\ 
  \hline
\textbf{Dual} & Exponential & 1 & 0.1 & \textbf{0.20} & \textbf{31.8\% (0.4)} \\ 
  High & Exponential & 1 & 0.1 & 0.23 & 34.0\% (0.5) \\ 
  Naive & Exponential & 1 & 0.1 & 0.30 & 43.6\% (0.4) \\ 
  Std & Exponential & 1 & 0.1 & 0.47 & 43.1\% (0.6) \\ 
  \textbf{Dual} & Exponential & 1 & 0.2 & \textbf{0.17} & \textbf{29.3\% (0.4)} \\ 
  High & Exponential & 1 & 0.2 & 0.20 & 31.0\% (0.4) \\ 
  Naive & Exponential & 1 & 0.2 & 0.29 & 43.0\% (0.3) \\ 
  Std & Exponential & 1 & 0.2 & 0.43 & 40.6\% (0.4) \\
  \hline
  \textbf{Dual} & Exponential & 2 & 0.1 & \textbf{0.18} & \textbf{30.6\% (0.4)} \\ 
  High & Exponential & 2 & 0.1 & 0.23 & 34.0\% (0.5) \\ 
  Naive & Exponential & 2 & 0.1 & 0.29 & 42.7\% (0.4) \\ 
  Std & Exponential & 2 & 0.1 & 0.43 & 40.6\% (0.4) \\ 
  \textbf{Dual} & Exponential & 2 & 0.2 & \textbf{0.15} & \textbf{28.5\% (0.3)} \\ 
  High & Exponential & 2 & 0.2 & 0.20 & 31.0\% (0.4) \\ 
  Naive & Exponential & 2 & 0.2 & 0.29 & 42.4\% (0.3) \\ 
  Std & Exponential & 2 & 0.2 & 0.41 & 40.5\% (0.3) \\
  \hline
  \textbf{Dual} & Exponential & 4 & 0.1 & \textbf{0.16} & \textbf{29.5\% (0.3)} \\ 
  High & Exponential & 4 & 0.1 & 0.23 & 34.0\% (0.5) \\ 
  Naive & Exponential & 4 & 0.1 & 0.29 & 42.3\% (0.3) \\ 
  Std & Exponential & 4 & 0.1 & 0.41 & 40.5\% (0.3) \\ 
  \textbf{Dual} & Exponential & 4 & 0.2 & \textbf{0.14} & \textbf{27.9\% (0.3)} \\ 
  High & Exponential & 4 & 0.2 & 0.20 & 31.0\% (0.4) \\ 
  Naive & Exponential & 4 & 0.2 & 0.28 & 42.1\% (0.3) \\ 
  Std & Exponential & 4 & 0.2 & 0.40 & 40.8\% (0.3)
\end{tabular}
\caption{Results for estimation and inference
  quality in 2D simulations when background signal has an Exponential
  correlation structure. As in Table \ref{tab:2d-sims-main}, results
  for the \emph{High} resolution method do not change across the
  different SNR ratios, but are 
  repeated to facilitate comparison. \emph{MSE} refers to mean squared
  error computed over the entire high resolution mean parameter
  vector. \emph{False --} reports the mean (SE)
  false negative error rate when the number of discoveries was fixed
  at 450. One hundred replicates per parameter combination.
  \protect{\label{supp:tab:2d-sims-exponential}}}
\end{table}

\begin{table}[ht]
\centering
\begin{tabular}{ l l c c c l }
  \multicolumn{1}{c}{\emph{Model}} 
  &  \multicolumn{1}{c}{\emph{Kernel}} 
  & \emph{SNR}$_s$\emph{:SNR}$_h$
  & \emph{SNR}$_h$
  & \emph{MSE}
  & \multicolumn{1}{c}{\emph{False --}} \\ 
  \hline
\textbf{Dual} & Gaussian & 1 & 0.1 & \textbf{0.24} & \textbf{29.8\% (0.3)} \\ 
  High & Gaussian & 1 & 0.1 & 0.28 & 34.1\% (0.3) \\ 
  Naive & Gaussian & 1 & 0.1 & 0.25 & 34.5\% (0.3) \\ 
  Std & Gaussian & 1 & 0.1 & 0.59 & 50.0\% (0.4) \\ 
  \textbf{Dual} & Gaussian & 1 & 0.2 & \textbf{0.17} & \textbf{24.2\% (0.2)} \\ 
  High & Gaussian & 1 & 0.2 & 0.21 & 27.1\% (0.2) \\ 
  Naive & Gaussian & 1 & 0.2 & 0.19 & 27.1\% (0.2) \\ 
  Std & Gaussian & 1 & 0.2 & 0.58 & 43.5\% (0.3) \\
  \hline
  \textbf{Dual} & Gaussian & 2 & 0.1 & \textbf{0.21} & \textbf{27.5\% (0.2)} \\ 
  High & Gaussian & 2 & 0.1 & 0.28 & 34.1\% (0.3) \\ 
  Naive & Gaussian & 2 & 0.1 & 0.24 & 33.3\% (0.3) \\ 
  Std & Gaussian & 2 & 0.1 & 0.58 & 43.5\% (0.3) \\ 
  \textbf{Dual} & Gaussian & 2 & 0.2 & \textbf{0.15} & \textbf{22.8\% (0.2)} \\ 
  High & Gaussian & 2 & 0.2 & 0.21 & 27.1\% (0.2) \\ 
  Naive & Gaussian & 2 & 0.2 & 0.18 & 26.4\% (0.2) \\ 
  Std & Gaussian & 2 & 0.2 & 0.57 & 38.6\% (0.2) \\
  \hline
  \textbf{Dual} & Gaussian & 4 & 0.1 & \textbf{0.18} & \textbf{25.0\% (0.2)} \\ 
  High & Gaussian & 4 & 0.1 & 0.28 & 34.1\% (0.3) \\ 
  Naive & Gaussian & 4 & 0.1 & 0.24 & 33.0\% (0.3) \\ 
  Std & Gaussian & 4 & 0.1 & 0.57 & 38.6\% (0.2) \\ 
  \textbf{Dual} & Gaussian & 4 & 0.2 & \textbf{0.12} & \textbf{21.1\% (0.2)} \\ 
  High & Gaussian & 4 & 0.2 & 0.21 & 27.1\% (0.2) \\ 
  Naive & Gaussian & 4 & 0.2 & 0.18 & 25.9\% (0.2) \\ 
  Std & Gaussian & 4 & 0.2 & 0.54 & 34.8\% (0.2)
\end{tabular}
\caption{Results for estimation and inference
  quality in 2D simulations when background signal has a Gaussian
  correlation structure. As in Tables \ref{tab:2d-sims-main} and
  \ref{supp:tab:2d-sims-exponential}, results 
  for the \emph{High} resolution method do not change across the
  different SNR ratios, but are 
  repeated to facilitate comparison. \emph{MSE} refers to mean squared
  error computed over the entire high resolution mean parameter
  vector. \emph{False --} reports the mean (SE)
  false negative error rate when the number of discoveries was fixed
  at 450. One hundred replicates per parameter combination.
  \protect{\label{supp:tab:2d-sims-gaussian}}}
\end{table}

\begin{supplement}
  \stitle{Covariance function estimation}
  \sdescription{
    In this section we detail our procedure to estimate isotropic
    covariance functions from 3D data; in practice
    the method could be extended to arbitrary $n$ dimensional data
    sources. The methods considered herein are not new but are included
    for completeness. We also report simulation
    results using this method to estimate the covariance from small three
    dimensional images and show that the method has relatively small bias
    in most simulation settings. }
  \label{supp:sec:mce}
\end{supplement}

\begin{algorithm}[H]
\caption{Minimum contrast estimation of $\btheta$: high level overview 
  \protect{\label{supp:alg:mce}}} 
\begin{algorithmic}[1]
\Procedure {EstimateImageCovariance}{$\bY$, $\mathrm{k}(\cdot;
  \btheta)$, 
  $\bTheta$} \Comment{With the argument to $\mathrm{k}(\cdot;
  \btheta)$ the 
  Euclidean distance between any two points, $\lVert \vox - \vox' 
  \rVert$} 
\State Inputs: Image $\bY$; covariance function $\mathrm{k}(\cdot; 
\bm{\theta})$ parameterized by $\bm{\theta}$ with feasible region
$\bTheta$ 
\State Construct $\mathcal{D} \gets
\textsc{ExtractCovarianceSummary}(\bY)$ \Comment{With $\mathcal{D} =
  (\bd, \hat{\bc}, \bm{\omega})$} 
\State Return $\argmin_{\btheta \in \bTheta} \sum_{i = 1}^{\text{dim}
  (\hat{\bc})} \omega_i [\hat{c}_i - \mathrm{k}(d_i; \btheta)]^2$ 
\EndProcedure
\end{algorithmic}
\end{algorithm}

Algorithm \ref{supp:alg:mce} outlines our minimum contrast estimation
(MCE) procedure at a high level. The algorithm first extracts
summary data $\mathcal{D} = (\bd, \hat{\bc}, \bm{\omega})$ from the
input data source $\bY$, where $\hat{\bc}$ are
empirical covariances between elements of $\bY$ offset by
corresponding distances $\bd$, and $\bm{\omega}$ is a set of
corresponding weights (defined below in algorithm
\ref{supp:alg:mce-summary-data}). The algorithm then finds $\btheta$
from within constraint region $\bTheta$ to minimize a weighted least
squares contrast between the $\hat{c}_i$ and $\mathrm{k}(d_i; \btheta)$.

With $\mathrm{k}(\cdot; \btheta)$ taken to be the radial basis function as in
\eqref{eqn:rbf}, for example, the parameters $\btheta$ correspond to the
marginal variance $\tau^2$, correlation bandwidth $\psi$, and exponent
$\nu$. For this problem, we took the feasible region $\bTheta$ to
constrain $0 < \tau^2 < \hat{c}_0$, $0 < \psi$, and $0 < \nu \leq 2$,
where $\hat{c}_0$ is the empirical variance of $\bY$. For problems we
consider, we found that the additional constraint $\psi \leq \nu$
frequently helped improve estimation.

Construction of $\mathcal{D}$ using a modified 3D
raster scan is outlined in algorithm
\ref{supp:alg:mce-summary-data}. In the algorithm, empirical
covariances between voxels and their neighbors are computed by
shifting the $(i, j, k)$ index of each voxel by the rows of the matrix
$\bP$ (which is constructed with the procedure outlined in Algorithm 
\ref{supp:alg:mce-perturbations}). The rows of $\bP$ define a series of
perturbations in a dense 3D raster scan. In one
dimension, a raster scan might only look ahead one pixel at a time so
as to visit each pair of adjacent pixels only once. In two dimensions,
the procedure might be defined to look ahead one pixel and look down
one pixel for the same reason. In three dimensions, a simple raster
might look ahead, down, and to the right by one or more voxels.
We designed our procedure to sample local pairs of voxels more densely
than this while still only visiting each unique pair once. Briefly, 
our algorithm ``looks ahead'' by visiting pairs of voxels within an
$(n_0 \times n_0 \times n_0)$ voxel cube such that the polar and
azimuthal angles of the search are between $[0^\circ, 180^\circ)$. We
further extended this search by adding simple raster scan
perturbations out to an $n_1$ voxel distance. In algorithm
\ref{supp:alg:mce-perturbations}, we defined $n_0 = 18$ voxels and $n_1
= 25$ voxels by default. Our default values encompass a large number
of perturbations while limiting the total computation time to a few
seconds for full scale brain images.

\begin{algorithm}[H]
\caption{Compute empirical covariance summary data
  \protect{\label{supp:alg:mce-summary-data}}} 
\begin{algorithmic}[1]
\Procedure {ExtractCovarianceSummary}{$\bY$, $n_0$, $n_1$}
\State Inputs: Image $\bY$ with dimensions $\bq \in \mathbb{R}^3$. We
set $n_0 = 18$, and $n_1 = 25$ by default 
\State Set $N \gets q_1 \cdot q_2 \cdot q_3$
\Comment{$N$ is the total number of voxels in image $\bY$}
\State Store $\bP \gets \textsc{ImageScanPerturbations}(n_0, n_1)$
\Comment{$\bP$ is an $(M \times 3)$ matrix of integers} 
\State Allocate $\bd \in \mathbb{R}^M$, $\hat{\bc} \in \mathbb{R}^M$
\Comment{$\bd$---perturbation distances; $\hat{\bc}$---empirical
  covariances} 
\State $\bs^{ab} \gets \bm{0}_M$, $\bs^{a} \gets \bm{0}_M$, $\bs^{b}
\gets \bm{0}_M$ \Comment{Accumulators for sufficient statistics} 
\State $\br \gets \bm{0}_M$ \Comment{Accumulators for counts of voxel
  pairs} 
\State Compute sufficient statistics for pairs of voxels separated by
perturbation distances: 
\For{$h$ in $1, \ldots, N$} \Comment{Outer loop over voxels} 
    \State Locate grid position $(i, j, k)$ such that corresponds to
    voxel $\vox_h$ 
    \If{$Y_{ijk}$ corresponds to brain data} 
        \For{$m$ in $1, \ldots, M$} \Comment{Inner loop over
          perturbations} 
            \State $(i', j', k') \gets (i, j, k) + \bP_m\trans$ 
            \If{$Y_{i'j'k'}$ corresponds to brain data}
            \Comment{Update sufficient statistics} 
                \State $s^{ab}_m \gets s^{ab}_m + Y_{ijk} \cdot
                Y_{i'j'k'}$ 
                \State $s^{a}_m \gets s^{a}_m + Y_{ijk}$; $s^{b}_m
                \gets s^{b}_m + Y_{i'j'k'}$ 
                \State $r_m \gets r_m + 1$ 
            \EndIf
        \EndFor
    \EndIf
\EndFor
\State Compute distances and empirical covariances associated with
grid perturbations: 
\State ``Locate'' voxel $\vox_0$ associated with grid position $(i, j,
k) = \bm{0}_3$ 
\For{$m$ in $1, \ldots, M$}
    \State ``Locate'' voxel $\vox'$ associated with grid position
    $\bP_m$ 
    \State $d_m \gets \lVert \vox_0 - \vox' \rVert$
    \If{$r_m > 1$}
        \State $\hat{c}_m \gets (s^{ab}_m - s^{a}_m s^{b}_m / r_m) /
        (r_m - 1)$ 
    \EndIf
\EndFor
\State Set $\omega_m \gets \#(\bd = d_m)$ \textbf{for} $m$ in $1,
\ldots, M$ \Comment{Count of instances of unique elements in $\bd$} 
\State Set $\omega_m \gets 1 / \omega_m$ \textbf{if} $\omega_m > 0$
\textbf{and} $\omega_m \gets 0$ otherwise \textbf{for} $m$ in $1,
\ldots, M$ 
\State Return $\mathcal{D} = (\bd, \hat{\bc}, \bm{\omega})$ 
\EndProcedure
\end{algorithmic}
\end{algorithm}

\begin{algorithm}[H]
\caption{Construct matrix $\bP$ of grid index perturbations for
  minimum contrast estimation procedure. 
  \protect{\label{supp:alg:mce-perturbations}}}
\begin{algorithmic}[1]
\Procedure {ImageScanPerturbations}{$n_0$, $n_1$}
\State Inputs: positive integers $n_0$, $n_1$, $n_0 < n_1$
\State Construct principal direction matrix $\bU \in \mathbb{R}^{(14
  \times 3)}$ such that each element $U_{ij} \in \{-1, 0, 1\}$; the
polar angle of each row of $\bU$ is between $[0^\circ, 180^\circ)$;
and the azimuthal angle of each row of $\bU$ is between $[0^\circ,
180^\circ)$. In our construction, $\bU$ includes a row of all 0's 
\State Construct $\bQ \in \mathbb{R}^{(n_0^3 \times 3)}$ with rows 
consisting of all 3-element permutations of elements of $(1, \ldots,
n_0)$ 
\State Compute $\bP \gets \bQ * \bU$, where $*$ denotes the
column-wise Khatri-Rao product 
\For{$k$ in $n_0 + 1, \ldots, n_1$}
    \State $\bP \gets [\bP\trans ~~ k\, \bI_3]\trans$
\EndFor
\State Remove duplicate rows from $\bP$
\State Return $\bP$
\EndProcedure
\end{algorithmic}
\end{algorithm}

\begin{figure}[!htb]
\centering
\includegraphics[width=0.7\textwidth]{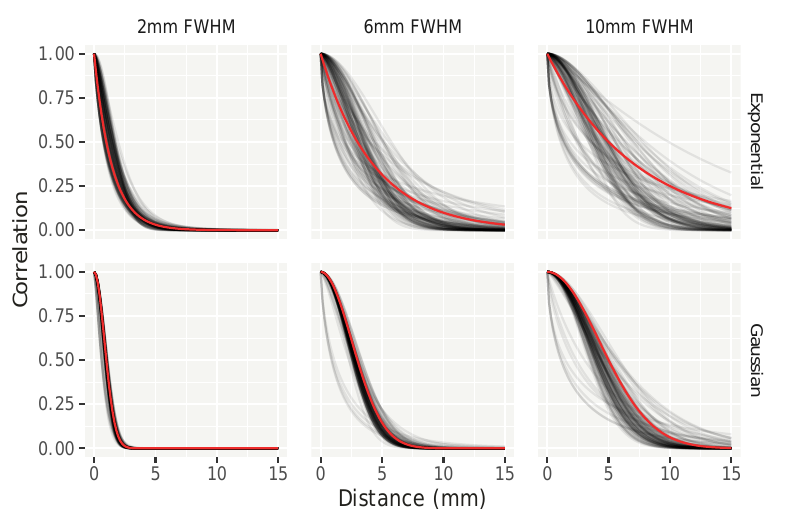}
\begin{tabular}{ l c c c }
\multicolumn{1}{c}{\emph{Kernel}} & \emph{FWHM} & \emph{Bias} & \emph{Variance} \\ 
  \hline
Exponential & 2 & $-6.73 \times 10^{-3}$ & $8.46 \times 10^{-4}$ \\ 
  Exponential & 6 & $-6.04 \times 10^{-2}$ & $7.22 \times 10^{-3}$ \\ 
  Exponential & 10 & $-1.47 \times 10^{-1}$ & $1.56 \times 10^{-2}$ \\ 
  Gaussian & 2 & $6.70 \times 10^{-3}$ & $1.17 \times 10^{-3}$ \\ 
  Gaussian & 6 & $-3.86 \times 10^{-3}$ & $3.45 \times 10^{-3}$ \\ 
  Gaussian & 10 & $-3.07 \times 10^{-2}$ & $1.73 \times 10^{-2}$
\end{tabular}
\caption{Recovery of the correlation function in small 3D
  images. Each gray line shows a correlation function estimated in
  repeated simulation (true correlation functions for each panel shown
  in red). In the table, \emph{Bias} and \emph{Variance} were computed
  pointwise and averaged over a dense grid from $[0, 15]$ (mm).
  \protect{\label{supp:fig:mce-simulations}}} 
\end{figure}

Fig. \ref{supp:fig:mce-simulations} presents the results of a
simulation assessing the performance of our MCE procedure. We
simulated small 3D images on a $(32 \times 32 \times
16)$ grid, treating voxels as isotropic 1 mm$^3$. In our simulation,
we drew mean images from Gaussian processes with either Exponential or
Gaussian correlation functions; unit marginal variance; and either
two, six, or ten mm full widths at half maxima. Mean images were
corrupted with independent Gaussian noise with the signal to noise
ratio set to 0.2 roughly to match our observed patient data and our 2D
simulations in section \ref{sec:simulations}. 

Since the Gaussian predictive process basis described in section
\ref{subsec:methods:prior-approximation} relies only on the
correlation bandwidth and exponent parameters $\psi$ and $\nu$, the
most important measure of estimation success in our setting is
recovery of the correlation function, not necessarily estimation of
$\btheta$ itself. For any given dataset, the nonlinear least squares
objective in algorithm \ref{supp:alg:mce} might be multimodal in
$\btheta$, but this is relatively unimportant if the resulting
correlation functions at different modes behave similarly.

In Fig. \ref{supp:fig:mce-simulations}, the red line in each panel
shows the true correlation function used to generate underlying mean
images in simulation, and the corresponding gray lines show estimated
correlation functions from 100 repeated simulations. The table below
the panels summarizes pointwise bias and variance averaged over a grid
of 1,000 equally spaced points from [0, 15] (mm). In the worst case
scenario (10 mm FWHM Exponential correlation function), pointwise mean
squared error was on average only about $3.8 \times 10^{-2}$, and was
between $[1.1, 7.5] \times 10^{-2}$ for 95\% of points on the 
grid. Even with relatively small 3D
images, these results suggest our MCE procedure can recover the
true correlation functions reasonably well.

\begin{supplement}
\stitle{Symmetry of our custom covariance function}
\sdescription{A remark on the cross covariance between
    $\mu(B_h)$ and $\mu(B_s)$ in our dual resolution mapping prior.}
\label{supp:sec:kernel-remark}
\end{supplement}

In the main body text, we defined a custom covariance function to
help map between high and standard spatial resolution images. We
reproduce that covariance function here for convenience:
\begin{equation*}
  K(\vox, \vox') = \begin{cases}
    k(\vox, \vox') & \text{if } \vox' \in B_h \\
    w\trans(\vox) k(B_h, \vox') &
    \text{otherwise}, 
  \end{cases}
\end{equation*}
where $w(\cdot) \approx K(B_h, B_h)^{-1} k(B_h, \cdot)$ (equations
\eqref{eqn:kernel} and \eqref{eqn:rbf} in the main text).
In our application, we take $k(\cdot, \cdot)$ to be the isotropic
radial basis function, 
\begin{equation*}
  k(\vox, \vox') = \tau^2 \exp(-\psi \lVert \vox - \vox'
  \rVert_2^\nu), \qquad \tau^2, \psi > 0, \quad \nu \in (0, 2].
\end{equation*}

\begin{rem}
  Under our prior,
  $\cov\{ \mu(\vox_h), \mu(\vox_s) \} = k(\vox_h, \vox_s)$ for any
  pair of $\vox_h \in B_h$ and $\vox_s \in B_s$.
\end{rem}

\begin{proof}
  Notationally, it is most convenient to show this relationship when
  $w(\vox) = K(B_h, B_h)^{-1} k(B_h, \vox)$ exactly, though the method
  is still valid given our approximation in section
  \ref{subsec:methods:prior-approximation},
  equation \eqref{eqn:basis-weights}. Per the definition of
  $K(\cdot, \cdot)$,
  \begin{align*}
    \cov\{ \mu(\vox_h), \mu(\vox_s) \}
    &= w\trans(\vox_h) k(B_h, \vox_s) \\
    &= k\trans(B_h, \vox_h) K(B_h, B_h)^{-1} k(B_h, \vox_s).
  \end{align*}
  Let $\bm{d} = [\I(\vox_i = \vox_h)]_{\vox_i \in B_h}$.
  Since $K(B_h, B_h) \bm{d} = k(B_h, \vox_h)$ by definition, it
  follows that,
  \begin{align*}
    k\trans(B_h, \vox_h) K(B_h, B_h)^{-1} k(B_h, \vox_s)
    &= \bm{d}\trans k(B_h, \vox_s) \\
    &= k(\vox_h, \vox_s).
  \end{align*}
\end{proof}



\bibliographystyle{imsart-nameyear} 
\bibliography{references}       


\end{document}